\documentclass[acmsmall]{acmart}

\usepackage{graphicx}
\usepackage[normalem]{ulem}
\useunder{\uline}{\ul}{}
\usepackage{caption}
\usepackage{subcaption}
\usepackage{bbold}
\usepackage[utf8]{inputenc}

\AtBeginDocument{%
  \providecommand\BibTeX{{%
    \normalfont B\kern-0.5em{\scshape i\kern-0.25em b}\kern-0.8em\TeX}}}

\setcopyright{acmcopyright}
\copyrightyear{2022}
\acmYear{2022}
\acmDOI{XXXXXXX.XXXXXXX}

\acmJournal{JACM}
\acmVolume{1}
\acmNumber{1}
\acmArticle{1}
\acmMonth{1}

\begin{document}

\title{Introducing a Framework and a Decision Protocol to Calibrate Recommender Systems}

\author{Diego Corrêa da Silva}
\email{diego.correa@ufba.br}
\orcid{0000-0001-7132-1977}
\authornotemark[1]
\affiliation{%
  \institution{Federal University of Bahia}
  \streetaddress{Av. Milton Santos, s/n}
  \city{Salvador}
  \state{Bahia}
  \country{Brazil}
  \postcode{40170-110}
}
\author{Frederico Araújo Durão}
\email{fdurao@ufba.br}
\orcid{0000-0002-7766-6666}
\authornotemark[1]
\affiliation{%
  \institution{Federal University of Bahia}
  \streetaddress{Av. Milton Santos, s/n}
  \city{Salvador}
  \state{Bahia}
  \country{Brazil}
  \postcode{40170-110}
}

\renewcommand{\shortauthors}{Silva and Durão}

\begin{abstract}
Recommender Systems use the user’s profile to generate a recommendation list with unknown items to a target user. Although the primary goal of traditional recommendation systems is to deliver the most relevant items, such an effort unintentionally can cause collateral effects including low diversity and unbalanced genres or categories, benefiting particular groups of categories. This paper proposes an approach to create recommendation lists with a calibrated balance of genres, avoiding disproportion between the user's profile interests and the recommendation list. The calibrated recommendations consider concomitantly the relevance and the divergence between the genres distributions extracted from the user's preference and the recommendation list. The main claim is that calibration can contribute positively to generate fairer recommendations. In particular, we propose a new trade-off equation, which considers the users' bias to provide a recommendation list that seeks for the users' tendencies. Moreover, we propose a conceptual framework and a decision protocol to generate more than one thousand combinations of calibrated systems in order to find the best combination. We compare our approach against state-of-the-art approaches using multiple domain datasets, which are analyzed by rank and calibration metrics. The results indicate that the trade-off, which considers the users’ bias, produces positive effects on the precision and to the fairness, thus generating recommendation lists that respect the genre distribution and, through the decision protocol, we also found the best system for each dataset.
\end{abstract}



\maketitle

\section{Introduction}
\label{intro}
Usually, recommender systems seek to recommend the most relevant items according to the user's profile. The recommendation is provided as a list ordered by the most relevant items, i. e., the more relevant the more on top. However, the focus on just recommending relevant items brings some drawbacks such as low diversity of categories \citep{Kaminskas:2016}, unbalanced genres \citep{Abdollahpouri:2020}, popularity bias \citep{Abdollahpouri:2017} or miscalibration \citep{Steck:2018}. Therefore, in this paper we investigate the calibration recommendation context.

A calibrated recommender system seeks to create a recommendation list, following the proportion of each area (genres, classes, or tags) in the user's preferences. To this end, the system needs to understand all user's interests as a distribution. This distribution will be used as target to the recommendation list, which is designed to be as little divergent as possible. When the recommendation list distribution is highly divergent with user's preferences distribution, the state-of-the-art calls it miscalibration \citep{Steck:2018, Lin:2020, DASILVA2021115112}. The miscalibration could simply mean that the system fails to provide accurate personalized items together with the fair treatment of the classes in the user's preferences \citep{Abdollahpouri:2020, DASILVA2021115112}. Thus, a calibrated system creates a recommendation list considering relevant items and fair representation of the user's interest classes. The state-of-the-art defines calibration as a way to provide some degree of fairness in which all user's interests are respected \citep{Steck:2018, Abdollahpouri:2020, DASILVA2021115112}.

As an example of calibration, in the song context, a user's preferences composed of 60\% Samba, 30\% Rock and 10\% K-Pop will create a recommendation list with these proportions. However, it is expected some divergence between the proportions on the user's preferences and the user's recommendation list, in the example it can be 56\% Samba, 31\% Rock and 13\% K-Pop. As introduced, the lower the divergence the better the calibration. 

In an implementation point-of-view, the calibrated recommendations \citep{Steck:2018} are described as a post-processing, which rearranges the recommendation list from the recommender algorithm. The state-of-the-art (Section \ref{sec:related_work}) develops many implementations, each one focused on their own proposals with workflows assigned by the system specialist. Based on that, our first research claims is a conceptual framework. We use the pattern of pre-processing, processing, and post-processing, commonly used in the field \citep{Pitoura:2021}, to decompose the calibrated system in these three steps and twelve components (Section \ref{sec:calibrated_recommendation}). To test the framework reproducibility, we implement 1092 calibrated recommendation systems (Section \ref{sec:experimental_setup}). In this sense, our second proposal is based on experiments that implement many calibrated systems and need to decide which is the best one. To automatically decide and indicate the best system to each dataset, we introduce the decision protocol and its coefficients (Section \ref{sec:decision_protocol}). Our third claim is a new formula to control the degree of relevant items and divergence (between the user's preferences and the user's recommendation list).

From the claims, we will conduct the discussion following the \textit{research questions} (Section \ref{sec:results}): 

\begin{itemize}
    \item{\textbf{RQ1}) How to find the best calibrated system to each domain?}
    \item{\textbf{RQ2}) Do the calibrated systems have the same performance on the datasets? Are there differences among the systems behavior?}
    \item{\textbf{RQ3}) Is it possible to create a framework model for calibrated systems? What are the important components to these systems?}
    \item{\textbf{RQ4}) When the user's bias is considered in the calibration trade-off, is it possible to improve the system performance?}
\end{itemize}

\section{Calibrated Recommendations}
\label{sec:calibrated_recommendation}
In this section, we cover the first part of our claims, the conceptual framework to calibrated recommendation systems. The framework is inspired by other studies \citep{Steck:2018, Kaya:2019, Abdollahpouri:2020, DASILVA2021115112} that exploit singular points of the calibrated system. These works don't present a generalized description of the calibrated system. The lack of specifications to produce modular systems may lead the specialist to avoid exploring new possibilities and to improve the system.

In addition to the conceptual framework, a new equation is proposed to balance the list of recommendations with relevant and fair items. As we proceed in the section, we describe the framework and introduce the equations used in our experiments. 

\subsection{The Framework}
As introduced, we claim a conceptual framework to calibrated recommendations. The framework is systematically divided into stages and components, which is described in order.

\subsubsection{System steps}
The calibrated recommendation systems can be described in three major stages: pre-processing, processing and post-processing. Each one has its restricted function in the system. The final stage is the most important, because that's where the fairer list of recommendations is created. Each stage is outlined as follows:

\begin{enumerate}
    \item The \textbf{pre-processing} deals with data treatment. This first step receives a raw input data set and returns one that can be used by the calibrated recommendation system. The usable data have the requirements to be used during the process stage. In Section \ref{sec:sub:datasets}, we present the two datasets, showing their numbers of users, items and interactions, before and after the pre-processing;
    \item \textbf{Processing} is the stage at which the recommender algorithm is implemented. It is common for the recommendation systems to produce a recommendation list with the recommender output. However, in calibrated recommendation systems, the output of the recommendation algorithm serves as the input for post-processing. In this paper, we refer to the recommendation list created by the recommender algorithm as candidate items. In Section \ref{sec:sub:methodology}, we discuss the seven recommenders used in our experiments; 
    \item The \textbf{post-processing} is responsible for receiving the candidate items and using them to create the fair recommendation list. In particular, this stage is described along this section.
\end{enumerate}

\subsubsection{System Components}
The \textbf{pre-processing} step is divided into three components: \textbf{cleaning}, \textbf{filtering} and \textbf{modeling}. Our framework defines three core components to the pre-processing, due to the calibration requires a class/genre in addition to the user interactions (rating/likes). The cleaning component removes data that cannot be used by the calibrated system, such as missing and incorrect information. The filtering component blocks data that can provide noise to the system. In the end, the pre-processing models the data to be used in the processing step.

For instance, Table \ref{tab:item_set_examples} provides an example of songs modeled under our conceptual framework. The information used to calibrate will be the genre of the songs. Table \ref{tab:user_set} presents three users and their information. We only use users' preferences, i. e., the interactions with the recommendation system. It is possible to apply other users' information depending on the recommendation technique. However, our framework is ready to work in the collaborative filtering \citep{Steck:2018, DASILVA2021115112} and content-based filtering \citep{Starychfojtu:2020}. Table \ref{tab:user_preference_set_example} shows the preferences of three modeled users, with the user-item-rating composing one tuple. We introduce two formal definitions in these examples: first the set of classes $C$ and a single class $c$, and second is the rating (or score, weight, times played) represented as $w_{ui}$.

\begin{table}[h!]
	\centering
	\caption{Eight items and their genres as classes.}
	\label{tab:item_set_examples}
	\begin{tabular}{cc}
		\hline
		\multicolumn{1}{r}{\textbf{UID}} & \textbf{$c \in C$}      \\ 
		\hline
		
		\multicolumn{1}{r}{I-001}        & Pop$|$Rock                \\ 
		\multicolumn{1}{r}{I-002}        & Pop \\ 
		\multicolumn{1}{r}{I-003}        & Blues             \\ 
		\multicolumn{1}{r}{I-004}        & Samba                \\ 
		\multicolumn{1}{r}{I-005}        & Funk                  \\ 
		\multicolumn{1}{r}{I-006}        & K-Pop             \\ 
		\multicolumn{1}{r}{I-007}        & MPB$|$Funk                 \\ 
		\multicolumn{1}{r}{I-008}        & Pop$|$Rock$|$Pagode$|$Funk       \\ 
		
		\hline
	\end{tabular}\\
	\centering \footnotesize{Source: Table created by the author.}
\end{table}

\begin{table}[h!]
	\centering
	\caption{Three hypothetical users.}
	\label{tab:user_set}
	\begin{tabular}{c}
		\hline
		\multicolumn{1}{c}{\textbf{UID}}                              \\ 
		\hline
		
		\multicolumn{1}{c}{U-001} \\ 
		\multicolumn{1}{c}{U-002} \\ 
		\multicolumn{1}{c}{U-003} \\
		
		\hline
	\end{tabular}\\
	\centering \footnotesize{Source: Table created by the author.}
\end{table}

\begin{table}[h!]
	\centering
	\caption{Three users' preference set.}
	\label{tab:user_preference_set_example}
	\begin{tabular}{ccc}
		\hline
		\multicolumn{1}{r}{\textbf{User UID}} & \multicolumn{1}{c}{\textbf{Item UID}} & \textbf{$w_{ui}$}      \\ 
		\hline
		
		\multicolumn{1}{c}{U-001}                                 & \multicolumn{1}{c}{I-001}                              & 1                                  \\ 
		\multicolumn{1}{c}{U-001}                                 & \multicolumn{1}{c}{I-002}                              & 4                                  \\ 
		\multicolumn{1}{c}{U-001}                                 & \multicolumn{1}{c}{I-008}                              & 5                                  \\ 
		\multicolumn{1}{c}{U-002}                                 & \multicolumn{1}{c}{I-004}                              & 2                                  \\ 
		\multicolumn{1}{c}{U-002}                                 & \multicolumn{1}{c}{I-007}                              & 10                                 \\ 
		\multicolumn{1}{c}{U-003}                                 & \multicolumn{1}{c}{I-001}                              & 9                                  \\ 
		\multicolumn{1}{c}{U-003}                                 & \multicolumn{1}{c}{I-003}                              & 11                                 \\ 
		\multicolumn{1}{c}{U-003}                                 & \multicolumn{1}{c}{I-004}                              & 3                                  \\ 
		
		\hline
	\end{tabular}\\
	\centering \footnotesize{Source: Table created by the author.}
\end{table}

The \textbf{processing} step is composed only by the \textbf{recommender algorithm} component, which is responsible for executing the recommender algorithm and producing the candidate items. In this second step any technique can be implemented as collaborative filtering, content-based filtering or hybrid technique, as well as its recommender algorithms. Most of the related works implement collaborative filtering. Thus, inspired by the state-of-the-art \citep{Steck:2018, Kaya:2019, DASILVA2021115112, Abdollahpouri:2020}, we implement this technique. In some of these works the recommenders provide a rank prediction \citep{Steck:2018, Kaya:2019} and others provide ratings predictions \citep{DASILVA2021115112}. Regardless of the recommender algorithm implementation the output of this step is the user's candidate items. Table \ref{tab:candidate_items_model} provides an example of three users and their candidate items. The formal representation of the predicted rating is $\hat{w}_{u,i}$. For each unknown item by the user, the recommender algorithm predicts a possible weight (e. g. rating) that the user will attribute to the item.

\begin{table}[h!]
	\centering
	\caption{Three hypothetical users and their candidate items.}
	\label{tab:candidate_items_model}
	\begin{tabular}{ccc}
		\hline
		\multicolumn{1}{r}{\textbf{User}} & \multicolumn{1}{c}{\textbf{Item}} & \textbf{$\hat{w}_{u,i}$}      \\ 
		\hline
		
		\multicolumn{1}{c}{U-001}                                 & \multicolumn{1}{c}{I-001}                              & 1                                  \\ 
		\multicolumn{1}{c}{U-001}                                 & \multicolumn{1}{c}{I-002}                              & 4                                  \\ 
		\multicolumn{1}{c}{U-001}                                 & \multicolumn{1}{c}{I-008}                              & 5                                  \\ 
		\multicolumn{1}{c}{U-002}                                 & \multicolumn{1}{c}{I-004}                              & 2                                  \\ 
		\multicolumn{1}{c}{U-002}                                 & \multicolumn{1}{c}{I-007}                              & 10                                 \\ 
		\multicolumn{1}{c}{U-003}                                 & \multicolumn{1}{c}{I-001}                              & 9                                  \\ 
		\multicolumn{1}{c}{U-003}                                 & \multicolumn{1}{c}{I-003}                              & 11                                 \\ 
		\multicolumn{1}{c}{U-003}                                 & \multicolumn{1}{c}{I-004}                              & 3                                  \\ 
		
		\hline
	\end{tabular}\\
	\centering \footnotesize{Source: Table created by the author.}
\end{table}

\textbf{Post-processing} is the most critical step for the calibrated recommendation system. It receives from the processing step a set of candidate items and uses them to create a final recommendation list. The objective of this step is to select a subset of items from the user's candidate items that are relevant and fair to the user. To achieve that many formulas are used to compose the workflow. Thus, the state-of-the-art \citep{Steck:2018, Kaya:2019, Abdollahpouri:2020, Starychfojtu:2020, DASILVA2021115112} implements different systems. Along this section, we investigate the basic component as well as some generalized equations.

\begin{figure*}[!ht]
	\centering
	\includegraphics[width=\linewidth]{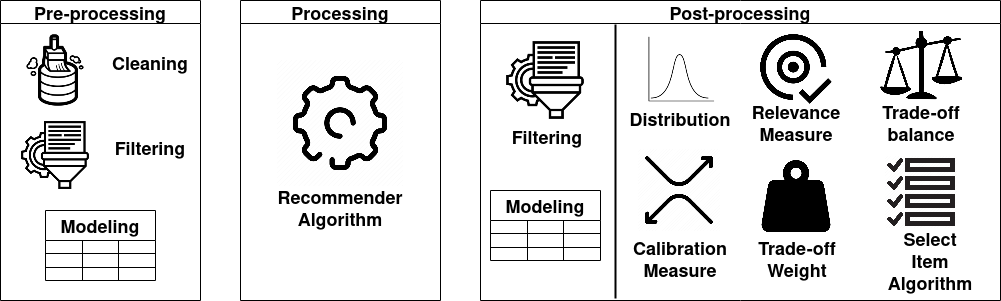}
	\caption{The conceptual framework with its steps and components.}
	\label{fig:framework}
\end{figure*}

The eight basic components of the post-processing are described as follows: i) filtering, ii) modeling, iii) distributions, iv) calibration measure (aka divergence measure), v) relevance measure, vi) trade-off weight ($\lambda$), vii) trade-off balance and viii) select item algorithm. The \textbf{filtering component}, on the post-processing, will filter the inputted candidate items. The state-of-the-art \citep{Kaya:2019, DASILVA2021115112, Steck:2018} uses this component to get the 100 most relevant candidate items or the 1000 top candidate items \citep{Seymen:2021}. In some cases, it is not possible to use all candidate items, due to the time-complexity which is a NP-hard problem \citep{Steck:2018, DASILVA2021115112}. Thus, inspired by the state-of-the-art we filter the 100 top relevant items in the candidate items set, i. e., the 100 items with high value in $\hat{w}_{u,i}$. The \textbf{modeling component} is responsible for moldeling the candidate items as a usable entrance to the post-processing. A predicted weight $\hat{w}_{u,i}$ is needed in each candidate item, thus the component prepares the items with the requested information and structure. All other components will be presented in sequence as sections. Figure \ref{fig:framework} shows the conceptual framework with its steps and components as described so far.

\subsection{Notations}
The mathematical notations used in the following sections are formally presented in Table \ref{tab:notation}. Along this section all notations are contextualized.

\begin{table}[hbt!]
    \centering
    \caption{Formal representation of the notations used in this study.}
    \label{tab:notation}
    \begin{tabular}{ll}
    \hline
    \textbf{Notation} & \textbf{Description}                                 \\ \hline
    $U$               & Set of all users                                     \\
    $u$               & A user                                               \\
    $I$               & Set of all items                                     \\
    $i$               & An item                                              \\
    $G$               & Set of all genres                                    \\
    $g$               & A genre                                              \\
    $I(u)$            & Set of items in the user preference $u$              \\
    $CI(u)$           & Set of candidate items to be recommended to user $u$ \\
    $L$               & A generic list with ordered items                    \\
    $W(i)$            & All feedbacks weights values over an item $i$        \\
    $w_{u,i}$         & Feedback weight value given by $u$ over $i$          \\
    $\hat{w}_{u,i}$   & Predicted feedback weight value for $u$ over $i$     \\
    $N$               & Length of the final recommended list                   \\
    $R^*$             & Final recommended list to a user                     \\ \hline
    \end{tabular}
	\centering \footnotesize{Source: Table created by the author.}
\end{table}

\subsection{Class Distributions}
As introduced, the calibration recommendation system aims to create a list of recommendations that tracks the distribution of user's preferences, i. e., the system extracts from the user's preferences items a target distribution which the recommendation list must achieve. Distributions are obtained from the item classes. Extracting is done in different ways by the state-of-the-art, some cases implements genres \cite{Steck:2018, DASILVA2021115112}, popularity \cite{Abdollahpouri:2021:POP} or sub-profiles \cite{Kaya:2019}.

\subsubsection{Class Approach}
Inspired by the state-of-the-art \cite{Steck:2018, DASILVA2021115112}, we consider as part of the distribution equation the genre probability of being chosen in an item. p(g$|$i) receives a genre ($g$) and an item ($i$) and verifies if the genre composes the item. If the genre belongs to the item, the equation returns values a value from the computation described as:

\begin{equation}
    \label{eq:probability_of_a_genre_in_an_item}
	p(g|i) = \frac{1}{|classesIn(i)|}.
\end{equation}

\subsubsection{Target Distribution}
It is based on the user's preferences items $I(u)$. $p$ is the formal representation of the Target Distribution \citep{Sacharidis:2019} and it is described as:

\begin{equation}
	p(g \text{\textbar} u) = \frac{\sum_{i \in I(u)} \mathbb{1}(g \in i) w_{u,i}\cdot p(g \text{\textbar} i)}{\sum_{i \in I(u)}\mathbb{1}(g \in i)w_{u,i}},
	\label{eq:preference_probability}
\end{equation}
	
\noindent where $w_{u,i}$ is the item relevance weight value (aka rating) from the user $u$ over an item $i$; $p(g \text{\textbar} i)$ is the genre probability value presented in Equation \ref{eq:probability_of_a_genre_in_an_item}; $\mathbb{1}(g \in i)$ indicates that if the genre $g$ is present in an item $i$, than the result returns 1, if it is not, the result returns 0.

\subsubsection{Realized Distribution}
It represents the distribution from the recommendation list $L$, which the system aims to produce in less divergence with $p$. $q$ is the formal representation of the Realized Distribution \citep{Sacharidis:2019} as follows:

\begin{equation}
	q(g \text{\textbar} u) = \frac{\sum_{i \in L(u)}\mathbb{1}(g \in i) \hat{w}_{u,i}\cdot p(g \text{\textbar} i)}{\sum_{i \in CI(u)}\mathbb{1}(g \in i) \hat{w}_{u,i}},
	\label{eq:genre_probability_recommendation}
\end{equation}

\noindent where $q$ is similar to $p$. The difference between them is the weighing strategy $\hat{w}_{u,i}$, which in our approach is the predicted relevance weight from the recommender algorithm.

Steck \cite{Steck:2018} proposes $\tilde{q}$ instead of $q$ to lead with zero values and it is presented as: 
\begin{equation}
    \tilde{q}(g \text{\textbar} u) = (1-\alpha) \cdot q(g \text{\textbar} u) + \alpha \cdot p(g \text{\textbar} u).
    \label{eq:balanced_realized_distribution}
\end{equation}

In the same sense of Steck, we use $\tilde{q} = q$.

\subsection{Calibration Measure}
\label{sec:divergence}
Likewise \cite{DASILVA2021115112}, in order to measure how much calibrated $p$ and $q$ are, we use three divergence calibration measures: 1) Kullback-Leibler ($F_{KL}$), 2) Hellinger ($F_{HE}$) and 3) Pearson Chi-Square ($F_{CHI}$). All measures work in an interval between $[0,\infty] \in \mathbb{R}$. High values indicate miscalibration between $p$ and $q$ and low values indicate calibration. In some works \citep{Zhao:2021, Seymen:2021}, the divergence measure is called calibration measure.

\subsubsection{Kullback-Leibler} 
The KL-divergence equation is presented as:

\begin{equation}
    F_{KL}(p,q) = \sum_{g \in G} p(g \text{\textbar} u)\log_{2}\frac{p(g \text{\textbar} u)}{\tilde{q}(g \text{\textbar} u)}.
    \label{eq:kl}
\end{equation}

The studies \cite{Steck:2018, Kaya:2019, DASILVA2021115112} implement the same equation.

\subsubsection{Hellinger}
Steck \cite{Steck:2018} alerts in his work that Hellinger can be used as a calibration measure. Consequently, Silva et al. \cite{DASILVA2021115112} implement it in their work as described below:

\begin{equation}
    F_{HE}(p,q) = \sqrt[]{2 \cdot \sum_{g \in G} (\sqrt[]{p(g \text{\textbar} u)} - \sqrt[]{q(g \text{\textbar} u)})^2}.
    \label{eq:he}
\end{equation}

\subsubsection{Pearson Chi Square} 
Silva et al. \cite{DASILVA2021115112} propose to use the Pearson Chi Square. By that we applied that measure too as follows:

\begin{equation}
    F_{CHI}(p,q) = \sum_{g \in G}\frac{(p(g \text{\textbar} u)-\tilde{q}(g \text{\textbar} u))^2}{\tilde{q}(g \text{\textbar} u)}.
    \label{eq:chi}
\end{equation}

\subsection{Relevance Rank Measure}
\label{sec:rank_sim}
Similarly to \cite{Steck:2018, Kaya:2019, DASILVA2021115112}, we use an equation to obtain a relevance rank weight as follows:

\begin{equation}
    Sim(L) = \sum_{i \in L}w_{r(u, i)},
    \label{eq:rank}
\end{equation}

\noindent where the $Sim(L)$ is the sum of the predicted relevance weight given by the recommender algorithm.

\subsection{Personalized Trade-Off Weight}
\label{sec:trade_off_weight}
Most of related works \citep{Steck:2018, Kaya:2019, Zhao:2020, Zhao:2021} use constant trade-off weights ($\lambda$). Silva et al. \cite{DASILVA2021115112} propose two personalized ways to find the trade-off weight. As proposed by our conceptual framework, the trade-off weight is a component of the post-processing step and such component acts as a parameter optimization. The state-of-the-art shows that the $\lambda$ produces changes on the recommendation list, affecting the system performance.

\subsubsection{Normalized Variance}
The equation calculates the distance along all values in the distribution, each of that value is a genre/class representation and the values can be between $[0,1] \in \mathbb{R}$. It is described as:

\begin{equation}
    mp(u) = \frac{\sum_{g \in G}p(g \text{\textbar} u)}{ \text{\textbar} G \text{\textbar} },
    \label{eq:mean_variance}
\end{equation}

\noindent where the result from $mp(u)$ is an average of the values in each genre. That value represents a center point among all genres.

\begin{equation}
    \lambda_{u} = 1 - \frac{\sum_{g \in G} \text{\textbar} p(g \text{\textbar} u) - mp(u) \text{\textbar} ^2}{ \text{\textbar} G \text{\textbar} },
    \label{eq:variance}
\end{equation}

\noindent where $\lambda_{u}$ is an average of the summation over all genres, computing the squared difference between a genre and the center point.

\subsubsection{Genre count}
A second equation to find the personalized trade-off weight is described as:

\begin{equation}
    \lambda_{u} = \frac{\sum_{g \in G}\mathbb{1}(g \in I(u))}{ \text{\textbar} G \text{\textbar} },
    \label{eq:count_class}
\end{equation}

\noindent where $\mathbb{1}(g \in I(u))$ returns $1$ if the genre $g$ is in an item of the user preference, otherwise $0$. After the summation, the average is computed and that final value indicates the degree of calibration/fairness with the user's preferences.

\subsection{Trade-Off Balance}
\label{sec:trade_off}
All equations presented so far are to compose the trade-off balance. We introduced ways to achieve the relevant and the fair portion of the trade-off balance, which brings all parties together to generate a fair list of recommendations. In our study, we implement the Steck's \cite{Steck:2018} claims, used by \cite{Kaya:2019, DASILVA2021115112} and propose a new trade-off balance.

The Maximum Marginal Relevance (\textit{MMR}) \citep{Carbonell:1998:UMD:290941.291025} is commonly used to maximize the choice of items in recommendation. It tends to get the optimal set of recommendation items $R^*$ from the set of candidate items $CI(u)$. The state-of-the-art \cite{Steck:2018, Kaya:2019, DASILVA2021115112} also uses the \textit{MMR}. The \textit{MMR} is used as the base to the trade-off balance and in our system $(\forall u \in U)$ is represented as: 

\begin{equation}
    R^* = \underset{ \text{\textbar} L \text{\textbar} =N, L \subset CI_{u}}{\arg\max} Trade(\lambda_u, L, p),
    \label{eq:mrr}
\end{equation}

\noindent where the $Trade(\lambda_u, L, p)$ is any trade-off balance. $L$ is a sub list from the candidate items, that list is an intermediate list used when building the final recommendation list $R^*$.

\subsubsection{Linear}
Steck \cite{Steck:2018} proposes a way to calibrate the recommendation list with a trade-off between relevance and calibration. In our study, we call it \textbf{linear} trade-off and is described as:

\begin{equation}
    TradeLin(\lambda_u, L, p) = (1-\lambda_u) \cdot Sim(L) - \lambda_u \cdot F(p,q(L)),
    \label{eq:linear_trade_off}
\end{equation}

\noindent where: $\lambda_u$ is any personalized trade-off weight from Section \ref{sec:trade_off_weight} (or constant weight to all users), Sim(L) is from Section \ref{sec:rank_sim} and F(p,q(L)) is any divergence/calibration measure from Section \ref{sec:divergence}.

\subsubsection{Logarithmic bias}
We claim a new trade-off balance to calibrate the recommendation list. Our balance considers the users' bias as a way to improve the calibration with the user tendencies, leading to a more specialized distribution without the super specialization problem. It is presented as:

\begin{equation}
    \label{eq:log_tradeoff}
    TradeLog(\lambda_u, L, p) = sign(TradeLin) \cdot \log^{abs(TradeLin) + 1} + \hat{b}_{u}(L),
\end{equation}

\noindent where $sign(TradeLin)$ is a function that returns the signal from the value obtained in the $TradeLin$ (Eq. \ref{eq:linear_trade_off}). The $\log^{abs(TradeLin) + 1}$ is a logarithm from the $TradeLin$ absolute value (i.e. casting to positive), due to the log proprieties do not comprehend negative values and the $+1$ is to avoid the log from returning zero. The $\hat{b}_{u}(L)$ is presented in sequence.

Koren and Bell \cite{Koren:2015} present a user and item bias equations. The item bias is described as: 

\begin{equation}
    \label{eq:item_bias}
    b_{i} = \frac{\sum_{u \in W(i)}(w_{u, i}-\mu)}{\alpha+ \text{\textbar} W(i) \text{\textbar} }.
\end{equation}

Our proposed users' bias is based on the items with the predicted relevance weight $w_{r(u, i)}$ in the place of relevance weight value $w_{u, i}$, this adaptation turns possible to apply the users' bias in the post-processing step. Our claim on the user's bias equation indicates how much the rating prediction by the recommender algorithm is close to the real ratings. The $\mu$ is an average of the relevance weights among all users' preferences. $\alpha$ and $\sigma$ values are to avoid division by zero and we adopt $0.01$ to both. The $\hat{b}_{u}(L)$ receives the list $L$ and finds the user bias for that list. The formal representation of the users' bias with the proposed adaptation is:

\begin{equation}
    \label{eq:user_bias_weight}
    \hat{b}_{u}(L) = \frac{\sum_{i \in L}(w_{r(u, i)}-\mu-b_i)}{\sigma+ \text{\textbar} L \text{\textbar} }.
\end{equation}

\subsection{Select items algorithm}
\label{sec:surrogate}

The last component is the select items algorithm. Here many re-ranking or selection algorithms can be applied. Most of related works implement the Surrogate Submodular \citep{Nemhauser:1978:AAM:3114181.3114410}, and similar to \cite{Steck:2018, Kaya:2019, DASILVA2021115112}, we also implement it. The Submodular algorithm workflow is:

\begin{itemize}
	\item $R^*$ starts with an empty list \{\};
	\item for each $N-1$ interaction with the candidate items, if an item $i$ tested as optimal in temporally list $L$ it is the one that maximizes $R^*$, then $i$ is appended into the list;
	\item in the end, the list has $N-1$ items and the last optimal item is appended, completing as well the top-n items $R^*$.
\end{itemize}
\section{Decision Protocol}
\label{sec:decision_protocol}
The main contribution of our work is a protocol for deciding which system is the best to be implemented driven by data domain. The decision protocol expects as an input two or more calibrated systems and, based on the coefficient results, it suggests the best one. These systems are combinations of calibrated one based on the proposed framework.

\subsection{Metrics}

The protocol uses three well-known metrics: The Mean Average Precision (MAP), a traditional metric used by the Recommendation Systems community; The Mean Average Calibration Error (MACE), and the Mean Rank MisCalibration (MRMC), proposed by \cite{DASILVA2021115112}. Each calibrated system needs to be evaluated by these metrics.

\subsection{Coefficients}

From those evaluation metrics values, we can obtain two coefficients: i) the Coefficient of Calibration Error (CCE) and ii) the Coefficient of MisCalibration (CMC). Both coefficients are part of our approach and determinant for the effectiveness of the protocol.

The Coefficient of Calibration Error (CCE) can be described as:
\begin{equation}
    CCE = \frac{\overline{MACE}}{\overline{MAP}}.
    \label{eq:coef_MACE_MAP}
\end{equation}

The Coefficient of MisCalibration (CMC) can be described as:
\begin{equation}
    CMC = \frac{\overline{MRMC}}{\overline{MAP}}.
    \label{eq:coef_MRMC_MAP}
\end{equation}

The proposal is a division between the MACE or MRMC \citep{DASILVA2021115112} by the MAP. All metrics evaluate the results of each position in the recommended list. In the Equations \ref{eq:coef_MACE_MAP} and \ref{eq:coef_MRMC_MAP}, the numerator is the MACE or MRMC results average and the denominator is the MAP results average, and all results are obtained over the same calibrated system. It's worth mentioning that when the MACE or MRMC is higher and the MAP is lower the obtained CCE or CMC value is higher, this means the system has more error or miscalibration than precision; and when the MACE or MRMC is lower and the MAP is higher the obtained CCE or CMC value is lower, which means the system has less errors (or miscalibration) and is more precise. In Equations \ref{eq:coef_MACE_MAP} and \ref{eq:coef_MRMC_MAP}, the variables $\overline{MACE}, \overline{MRMC}$ and $\overline{MAP}$ are representations of each metric average, computed over each system running. 

\subsection{The decision}

From the coefficients, the decision protocol indicates which system combination is the best to be implemented. At this point we have, for each calibrated system, two coefficient values. The protocol uses the coefficients in an addition as:

\begin{equation}
    s_i = CCE + CMC,
    \label{eq:system_performance}
\end{equation}

\noindent where the system $i$ has its performance ($s_i$) measured with the coefficients, representing its total of error and miscalibration over the precision as previously debated.

The final decision is obtained by: 

\begin{equation}
    S = \min(s_1,\ s_2,\ s_3...\ s_n),
    \label{eq:decision}
\end{equation}

\noindent where the $n$ calibrated systems are judged and the system with the lower value of $s$ is chosen as the best calibrated system $S$.

In the next sections we will describe the experimental setup and results. Our research implements more than one thousand calibrated systems (considering the parameter optimization) and at the end suggests the best implementation based on the proposed decision protocol.
\section{Related Work}
\label{sec:related_work}
In recent years, the calibrated recommendation has attracted attention as a means of achieving fairness. This topic of research addresses different points-of-view, for instance, some state-of-the-art analyses the context of calibration \citep{Abdollahpouri:2020, Lin:2020}, others focused on creating a calibrated recommendation list in a user-centric view \citep{Steck:2018, Kaya:2019, Abdollahpouri:2021:POP, DASILVA2021115112} and others seek to provide more fairness in an item-centric view \citep{Zhao:2020}. Our research focuses on the user-centered view, i.e., a C-fairness system \citep{DASILVA2021115112}. In a sequence, we discuss several related works that generate calibrated recommendation list by the post-processing.

Steck \citep{Steck:2018} proposes the calibrated recommendations as a way to provide fairness, in which the recommendation list is generated to be the least divergent with the movie genres in the user's profile. The author implements the collaborative filtering technique through a matrix factorization recommendation algorithm. This algorithm provides a list of recommendations that is post-processed to ensure a certain degree of fairness. Steck implements the calibrated system as a unique system divide in three steps. The results show the importance of calibrating recommendations where the objective is to obtain a certain degree of fairness, but the accuracy is slightly reduced. Inspired by this work \citep{Steck:2018}, we implement the linear trade-off, the relevant sum, the distribution, the surrogate submodular and the Kullback-Leibler. So Steck's work is covered by our findings. Furthermore, the author suggests that the Hellinger divergence measure may be implemented as fairness measures, consequently we follow the Steck suggestion. The work uses only movie genres and doesn't cover other domains as music or books.

In a similar fashion with Steck \cite{Steck:2018} and expanding the study about calibrated recommendations, Kaya and Bridge \cite{Kaya:2019} investigate if calibrated recommendations can improve diversification in the recommendation list. Beyond the calibrated systems, the authors implement the Intent-aware approach, seeking to compare the results. As the authors explain, the Intent-aware is an approach originally from the Information Retrieval (IR) and adapted to the Recommendation System. The approach works with ambiguous representation, for instance, the requisition for ``Jaguar'', the system can have multiple representations for the requisition as cats, cars or operating systems. By that the approach seeks to include the greatest number of representations as possible in the list. Thus, the study \cite{Kaya:2019} implements two Intent-aware adaptations to recommendation systems the Query Aspect Diversification (xQUAD) and Subprofile-Aware Diversification - (SPAD). The authors compare these two Intent-aware forms with the Steck's \cite{Steck:2018} calibrated system and extend the system adding a new way to find the user profile distribution, as well as comparing four systems. The authors evaluate the results in three perspectives: precision, diversity and miscalibration. As part of methodology the authors use: the Movielens 20M and the Taste Profile as datasets, different trade-off weight values from \cite{Steck:2018} and a fast alternative least-squares matrix factorization recommender. The results indicate a relation that intent-aware calibrates the recommendation list to some degree and calibrated recommendations somehow imply a diverse recommendation. To the Taste Profile dataset, the calibrated system with the distribution based on subprofile increases the precision when the trade-off weight is high. As Steck \cite{Steck:2018}, Kaya and Bridge \cite{Kaya:2019} results indicate that the higher the trade-off value, the more calibrated the list will be. From \cite{Kaya:2019}, we cover all implementations inspired by \cite{Steck:2018}. We do not cover diversification in our study.

Abdollahpouri et al. \cite{Abdollahpouri:2020} study a connection between popularity bias and calibration. The short paper is focused on analyzing the calibration aspect and its fairness. It is argued that the groups which receive popularity bias amplification will receive a higher miscalibration degree too. Beyond, it is discussed that if the popularity bias is fixed it is possible that the disparate miscalibration can be fixed too, turning the system fairer. To create the groups, the authors divide all users in 10 groups based on the degree of popular items each user likes. They do not implement a post-processing step; the connection is between the user's preferences set and the recommendations provided by five chosen recommender algorithms. Abdollahpouri et al. \cite{Abdollahpouri:2020} use a different genre distribution formulation without probability and rank weight, as proposed in previous calibrated work. As a dataset they use Movielens 1M and core-10 Yahoo Movie and evaluate with miscalibration metric proposed by \cite{Steck:2018} and algorithmic popularity lift proposed by the authors in previous work. The results indicate that the users with less popular items are affected by the popularity bias and receive more miscalibrated recommendations.

Lin et al. \cite{Lin:2020} analyses several categories of profile characteristics over many factors that contribute to miscalibration for some users but not others. These characteristics are provided by the users profiles and can be interpreted from the users' preference. The category described in \cite{Lin:2020} is a generic word to class, tag or a genre as in our work. To evaluate the categories authors use the miscalibration proposed by \cite{Steck:2018} as well as the normalized Discounted Cumulative Gain (NDCG). As datasets they use the Movielens 1M and Yelp.com. As \cite{Abdollahpouri:2020}, \cite{Lin:2020} do not implement the post-processing step and use five well-known algorithms. The results show that for all algorithms and for both datasets the miscalibration is present in different degrees, where they analyze the characteristics depending on it the miscalibration present at different degrees too.
 
Seymen et al. \cite{Seymen:2021} propose a new trade-off balance called Calib-Opt. As calibration measure, the authors use the proposed Weighted Total Variation. In the rank measure they consider two news constraints. The selector algorithm used by \cite{Steck:2018} called surrogate submodular is substituted by Gurobi branch\&bound. The results indicate that the proposed Calib-Opt obtained a better performance than \cite{Steck:2018}.

Silva et al. \cite{DASILVA2021115112} propose a set of calibrated recommendations systems and metrics. The authors use the divergence measure Kullback-labler similar to \cite{Steck:2018, Kaya:2019} and propose the Hellinger and Pearson Chi Square. To evaluate the systems \cite{DASILVA2021115112} propose two new metrics to the calibrated context, the Mean Average Calibration Error (MACE) and the Mean Rank MisCalibration (MRMC). To evaluate the authors use six well-known collaborative filtering recommender algorithms. The results indicate that the Pearson Chi Square can obtain a better performance than other divergence measures. 

Pitoura et al. \cite{Pitoura:2021} conduct an overview about fairness in ranking and recommendations. The authors argue that calibrated recommendations are one model of fairness, which has six in total: demographic parity, conditional parity, equalized odds, fairness through awareness, counterfactual fairness and calibration-based fairness. In the overview, the authors define five viewpoints on fairness: fairness for the recommended items, fairness for the users, fairness for groups of users, fairness for the item providers and the recommendation platform. The study considers that the post-processing is the core for fairness systems, due to be the step that ensures the fairness, modifying the recommender algorithm output. 

\section{Experimental Setup}
\label{sec:experimental_setup}

In this section, we illustrate the experimental setup applied in our study. We describe the used datasets and methodology. Our setup is modeled to provide a better understanding of the system behavior. The datasets and methodology are inspired by related works in the state of art \citep{Steck:2018, Kaya:2019, DASILVA2021115112}.

\subsection{Datasets}
\label{sec:sub:datasets}
Inspired by \cite{Kaya:2019}, we use two public datasets for evaluating the proposal. Table \ref{tab:datasets} shows a data description from both datasets.

\begin{table}[t!]
    \centering
    \caption{Datasets before and after the pre-processing.}
    \label{tab:datasets}
    \begin{tabular}{lcccc}
        \hline
        \textbf{Datasets}   & \textbf{|U|}       & \textbf{|I|}     & \textbf{|W|}        & \textbf{|G|} \\ \hline
        Raw ML20M           & 138,493   & 27,278  & 20,000,263 & 19  \\
        Used ML20M         & 80,672    & 15,400  & 9,013,655  & 19  \\
        Raw Taste Profile   & 1,019,318 & 999,056 & 48,373,586 & 16  \\
        Used Taste Profile & 94,611    & 98,305  & 4,807,615  & 16  \\ \hline
    \end{tabular}
\end{table}

For the movie domain, we use the MovieLens 20M dataset (ML20M)\footnote{https://www.kaggle.com/grouplens/movielens-20m-dataset}\citep{Harper:2015}, which is the same dataset used by \cite{Steck:2018, Kaya:2019, DASILVA2021115112}. Table \ref{tab:datasets} show the numbers of users (|U|), items (|I|), feedbacks (|W|) and genres (|G|) are presented. As defined in the proposed framework the pre-processing applies cleaning, filtering and modeling to the data. Based on that and inspired by \cite{Steck:2018,Kaya:2019, DASILVA2021115112}, we apply a rating cut by 4 and dropping all lower ratings, the removal of movies without genre information and the removal of movies without user interactions. Unlike \cite{Steck:2018,Kaya:2019, DASILVA2021115112}, we dropped users with a preference set size smaller than 30 items. In addition, we dropped items with less than 3 interactions with users to avoid the item cold-start.

For the song domain, we use the Taste Profile\footnote{http://millionsongdataset.com/tasteprofile/} dataset and the Tagtraum genre annotations\footnote{http://www.tagtraum.com/genres/msd\_tagtraum\_cd2.cls.zip}, which is the same dataset used by \cite{Kaya:2019}, though is not the same genre annotation. Table \ref{tab:datasets} shows the numbers of users, items, feedbacks/transactions and genres are presented. For obtaining pre-processed data as indicated in the framework and similarly to \cite{Kaya:2019} we applied: the elimination of songs without genre information. Differently from \cite{Kaya:2019}, we dropped users with a preference set size smaller than 30 items, whose value was defined is defined based on the size of the recommendation list and a cut by the most played songs where it discards the ones played less than 3 times. In addition, we dropped items with less than 3 interactions with users to avoid the item cold-start.

\subsection{Methodology}
\label{sec:sub:methodology}

With the pre-processed data we ran the Grid Search method. Each recommender algorithm used in our experimentation searches the best hyper parameters with a 3-fold cross validation methodology. In order to choose the best hyper parameters (HP), we use the Mean Average Error (MAE) as a decision metric, i.e., the HP combination with the smallest error is chosen. The MAE measures the absolute error between the real and predicted weight, thus the smaller the MAE error, the smaller the error propagation for post-processing. To better understand the effects of the post-processing and test our decision protocol, as well as our framework, we select seven well-known recommender algorithms as shown in the \textbf{processing step}. They are: 

\begin{enumerate}
	\item \textbf{Basic UserKNN}: User-based K Nearest Neighbors \citep{Koren:2010:FNS:1644873.1644874}. The best HP is: $k=30$ as the number of neighbors and Mean Squared Difference as similarity for both datasets;
	
	\item \textbf{Basic ItemKNN}: Item-based K Nearest Neighbors \citep{Koren:2010:FNS:1644873.1644874}. The best HP is: $k=30$ as the number of neighbors and Person Correlation as similarity for both datasets;
	
	\item \textbf{Slope One}: \cite{Lemire:2007} presents the algorithm which has no HP;
	
	\item \textbf{NMF}: Non-negative Matrix Factorization presented by \cite{Luo:2014}. The best HP are: i) for ML20M the number of epochs $ne=50$, number of latent factors $f=50$, user and item learning rate $\gamma_u=0.005$ and $\gamma_i=0.005$, user and item regularization constant $\lambda_u=0.005$ and $\lambda_i=0.05$, and user and item regularization bias $\lambda_{bu}=0.005$ and $\lambda_{bi}=0.005$; ii) for Taste Profile the number of epochs $ne=30$, number of latent factors $f=50$, user and item learning rate $\gamma_u=0.003$ and $\gamma_i=0.005$, user and item regularization constant $\lambda_u=0.003$ and $\lambda_i=0.05$, and user and item regularization bias $\lambda_{bu}=0.003$ and $\lambda_{bi}=0.005$;
    
    \item \textbf{SVD}: Singular Value Decomposition proposed by \cite{Koren:2009:MFT:1608565.1608614}. The best HP are: i) for ML20M the number of epochs $ne=50$, number of latent factors $f=50$, user and item learning rate $\gamma_u=0.005$ and $\gamma_i=0.005$, and user and item regularization $\lambda_u=0.01$ and $\lambda_i=0.01$; ii) for Taste Profile the number of epochs $ne=30$, number of latent factors $f=150$, user and item learning rate $\gamma_u=0.001$ and $\gamma_i=0.001$, and user and item regularization $\lambda_u=0.05$ and $\lambda_i=0.05$;
	
	\item \textbf{SVD++}: An extension of SVD presented by \cite{Koren:2008:FMN:1401890.1401944}. The best HP are: for ML20M and Taste Profile the number of epochs $ne=50$, number of latent factors $f=50$, user and item learning rate $\gamma_u=0.005$ and $\gamma_i=0.005$, and user and item regularization constant $\lambda_u=0.01$ and $\lambda_i=0.01$;
	
	\item \textbf{Co Clustering}: Presented by \cite{George:2005}, the Co Clustering is a recommender based on clusterization. The best HP is: i) for ML20M the number of epochs $ne=50$, number of user and item clusters $\overline{C_{u}}=3$ and $\overline{C_{i}}=7$; ii) for Taste Profile the number of epochs $ne=10$, number of user and item clusters $\overline{C_{u}}=3$ and $\overline{C_{i}}=3$.
\end{enumerate}

This study does not propose any change in the recommender algorithms workflow. Therefore, we use the Python Surprise library \citep{Surprise} implementation to perform our experiments. The selected collaborative filtering recommenders used in the experiment are presented in the literature by \citep{Abdollahpouri:2020, Lin:2020, DASILVA2021115112}. 

The datasets used in the experiments are randomly divided for each user in 70\% of training data and 30\% of test data. This methodology of validation is use by the-state-of-the-art \cite{Steck:2018, Kaya:2019, DASILVA2021115112}. All test data is mixed with all unknown items by the user and it is expected the test items to compose the final recommendation list. The training data are given as an input to the recommender in order to learn about the users' preference. Having the recommender totally trained, the algorithm predicts the rating to all unknown+test items for each user. After predicting all possible ratings to the users-item matrix, similarly to \cite{Steck:2018, Kaya:2019, DASILVA2021115112}, we get the top 100 candidate items with the highest rating predicted by the recommender algorithm. The post-processing to each user starts with the top-100 candidate items, from which the system will select the final recommendation list.

In addition to the two personalized trade-off weights, we use constant values of $\lambda$. These same weighing values are also used by \cite{Kaya:2019, DASILVA2021115112} and they are $\lambda \in [0.0; 0.1; ...; 0.9; 1.0]$.

All the system combinations are generated taking into account two datasets, seven recommender algorithms, two trade-off balances, two personalized weights (VAR, CGR) plus eleven constant values (totaling 13 trade-off weights), and the three divergence measures. Each configuration combines a different way to generate the final recommendation list, so in total we have $2\cdot7\cdot2\cdot13\cdot3=1092$ different combinations to be evaluated by each metric. We run each of that combinations 3 times with different training and test data.

The final recommendation list $R^*$ is generated with the top-10 fair items given by the post-processing. Finally, we evaluate the recommendation list in all $[1;10] \in \mathbb{N}$ positions.
\section{Experimental Results}
\label{sec:results}

In this section we present the experimental results based on the setup described previously. In particular, we present the cross-metric between MAP and MACE as well as CCE (Subsection \ref{sec:sub:map_mace}); the crossed results between MAP and MRMC, plus CMC (Subsection \ref{sec:sub:map_mrmc}); and the protocol decision (Subsection \ref{sec:sub:decision}). Section \ref{sec:sub:research_question} answers all the research questions presented in the paper introduction. Section \ref{sec:sub:discussion} discuss the overall findings.

\subsection{MAP x MACE}
\label{sec:sub:map_mace}
First we will analyze the results obtained from the MAP and MACE metrics. To do that, we combine the results of each system combination, where each dot in the lines is a $\lambda$. Figure \ref{fig:ml_map_mace} presents the cross-metric MAPxMACE for the Movielens, as well as Table \ref{tab:CCE_ML_MACE_MAP} presents the CCE values. Figure \ref{fig:oms_map_mace} presents the cross-metric for the Taste Profile, whereas Table \ref{tab:CCE_OMS_MACE_MAP} presents the CCE values. 

\subsubsection{Movielens}
\begin{table*}[t!]
    \centering
    
    \caption{Movielens - CCE - Results from $LIN$ and $LOG$ trade-offs combined with the KL, HE and CHI divergence measure for each recommender algorithm.}
    \label{tab:CCE_ML_MACE_MAP}
    \resizebox{\textwidth}{!}{%
    \begin{tabular}{c|c|ccccccc}
        \hline
        \textbf{Divergence} & \textbf{Trade-off} & \textbf{\textbf{Co Clustering}} & \textbf{\textbf{Item KNN}} & \textbf{\textbf{NMF}} & \textbf{\textbf{Slope One}} & \textbf{\textbf{SVD}} & \textbf{\textbf{SVD++}} & \textbf{\textbf{User KNN}} \\ \hline
        KL                  & $LIN$              & 60.44                           & 90.73                      & 3.8                   & 63.52                       & 4.95                  & {\ul \textbf{3.52}}     & 430.91                     \\
        KL                  & $LOG$              & 62.94                           & 110.46                     & 5.76                  & 83.46                       & 4.31                  & {\ul \textbf{3.05}}     & 130.58                     \\ \hline
        HE                  & $LIN$              & 34.7                            & 57.4                       & 2.51                  & 40.28                       & 3.68                  & {\ul \textbf{2.38}}     & 711.24                     \\
        HE                  & $LOG$              & 57.84                           & 157.32                     & 8.31                  & 104.41                      & 3.06                  & {\ul \textbf{2.08}}     & 60.73                      \\ \hline
        CHI                 & $LIN$              & 38.26                           & 63.33                      & {\ul \textbf{2.92}}   & 47.41                       & 4.29                  & 3.06                    & 364.44                     \\
        CHI                 & $LOG$              & 50.63                           & 88.66                      & 4.83                  & 71.43                       & 4.3                   & {\ul \textbf{3.06}}     & 120.71                     \\ \hline
    \end{tabular}%
    }
\end{table*}

\begin{figure*}[ht!]
	\centering
	\begin{subfigure}{0.48\textwidth}
		\centering
		\includegraphics[width=\linewidth]{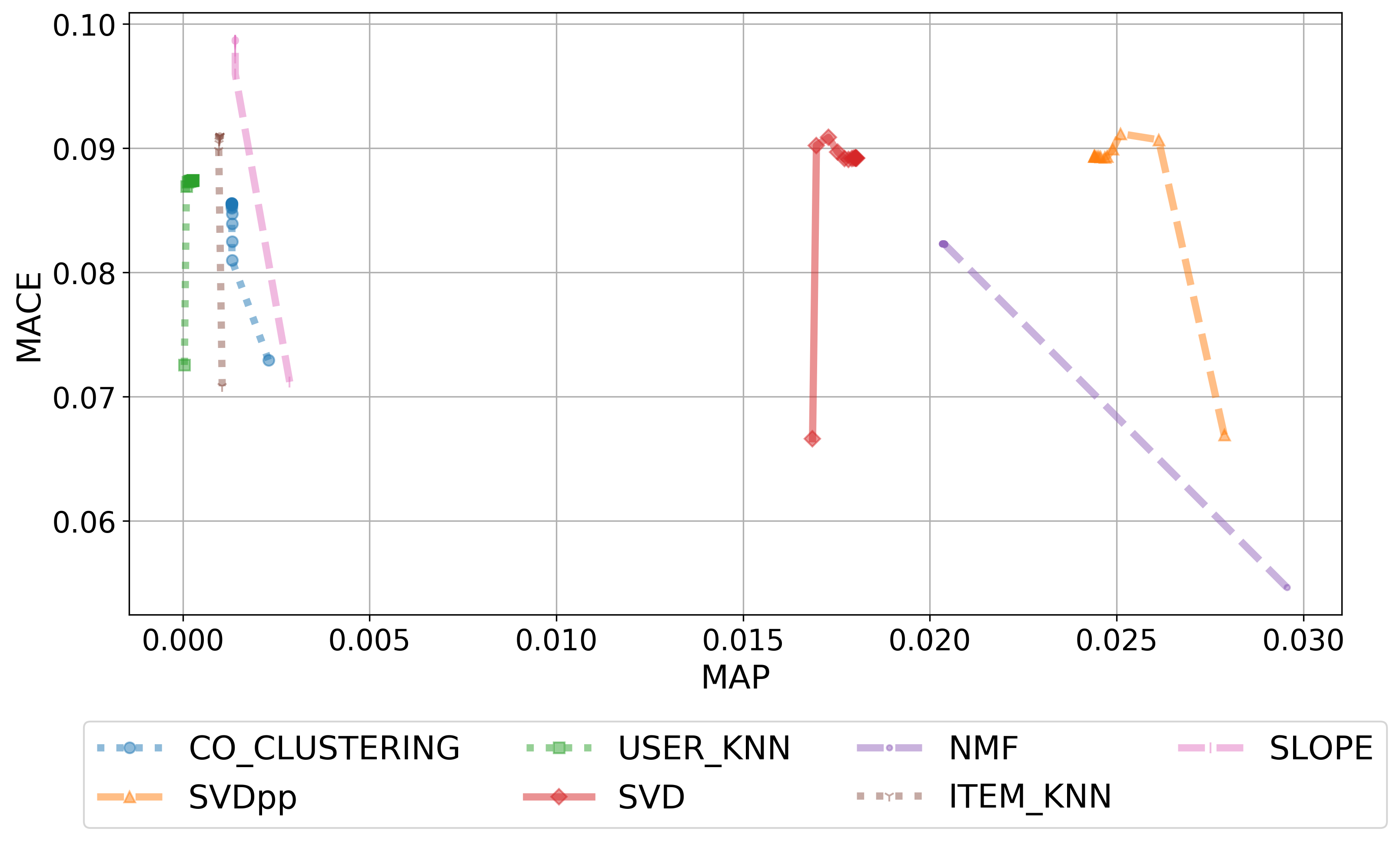}
		\caption{KL and $LIN$}
		\label{fig:ml_map_mace_kl_lin}
	\end{subfigure}
	\begin{subfigure}{0.48\textwidth}
		\centering
		\includegraphics[width=\linewidth]{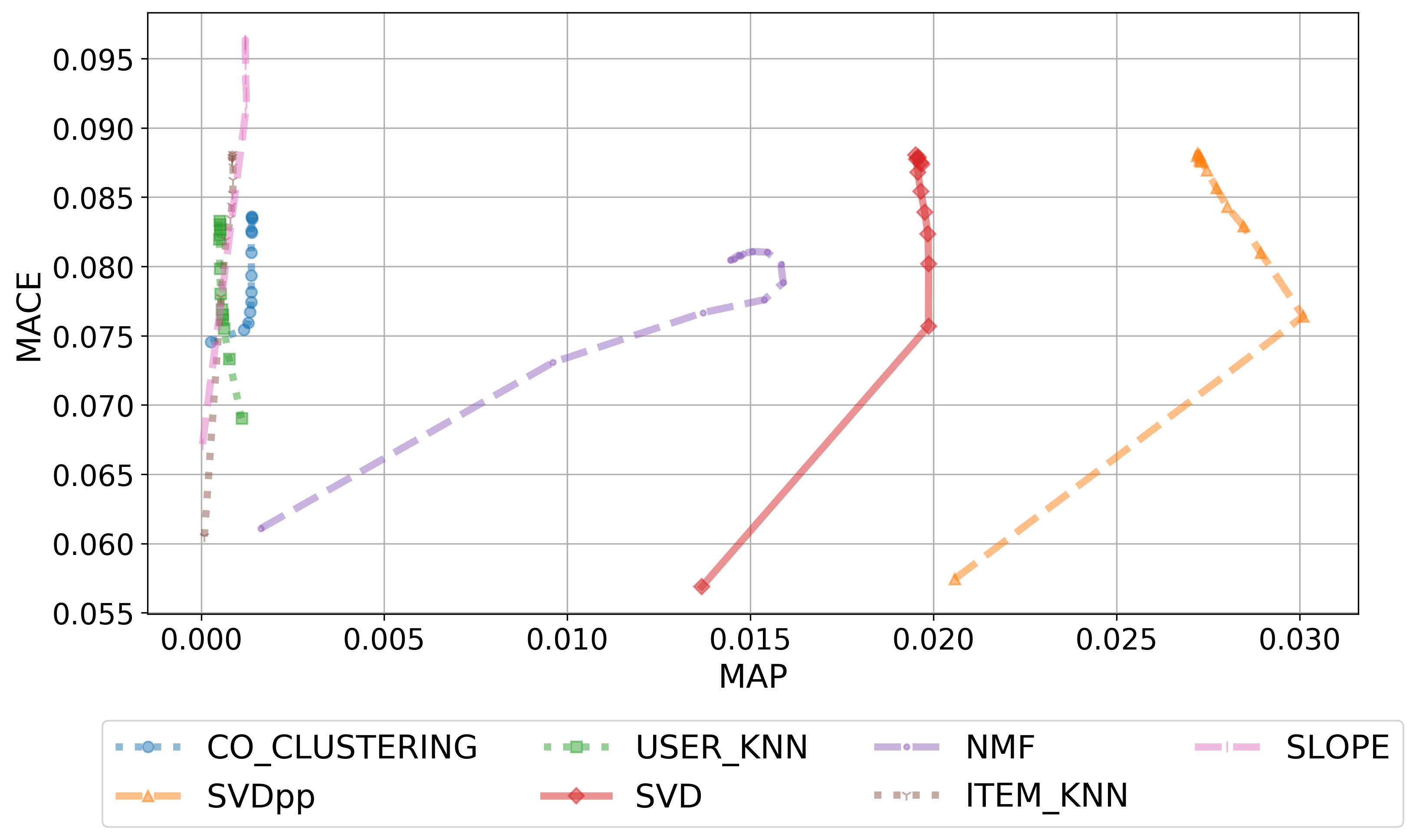}
		\caption{KL and $LOG$}
		\label{fig:ml_map_mace_kl_log}
	\end{subfigure}
	
	~
	
	\begin{subfigure}{0.48\textwidth}
		\centering
		\includegraphics[width=\linewidth]{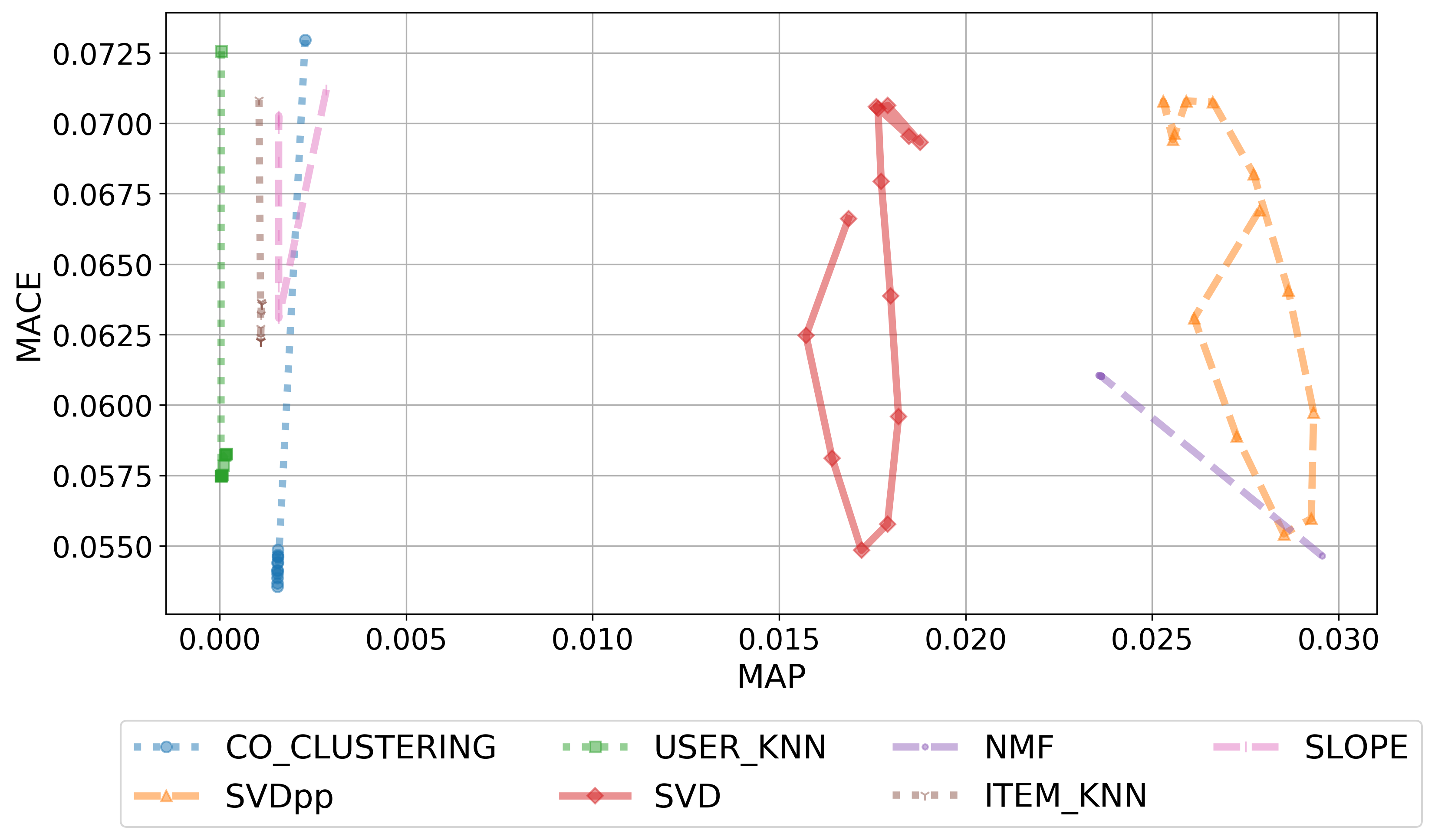}
		\caption{Hellinger and $LIN$}
		\label{fig:ml_map_mace_he_lin}
	\end{subfigure}
	\begin{subfigure}{0.48\textwidth}
		\centering
		\includegraphics[width=\linewidth]{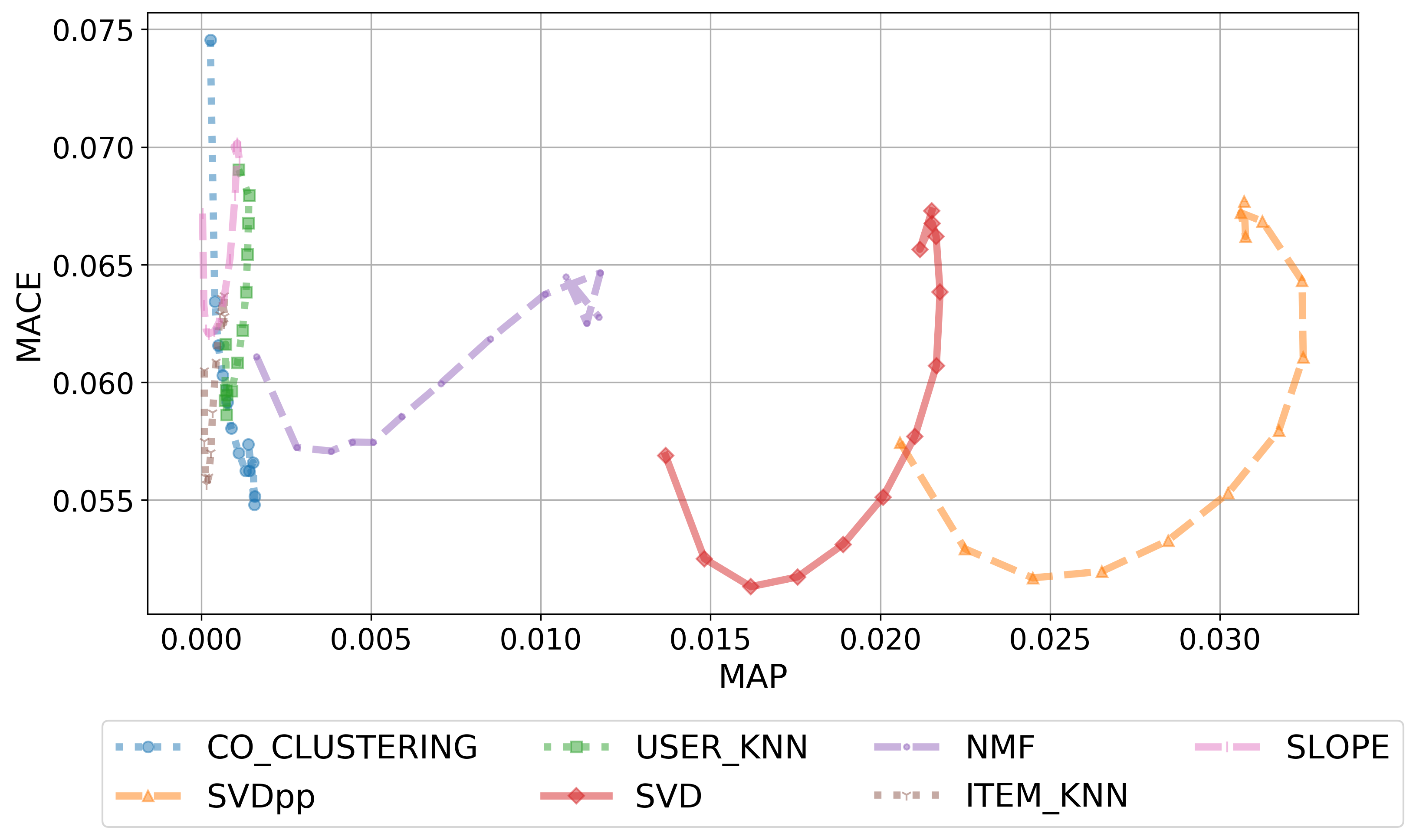}
		\caption{Hellinger and $LOG$}
		\label{fig:ml_map_mace_he_log}
	\end{subfigure}
	
	~

	\begin{subfigure}{0.48\textwidth}
		\centering
		\includegraphics[width=\linewidth]{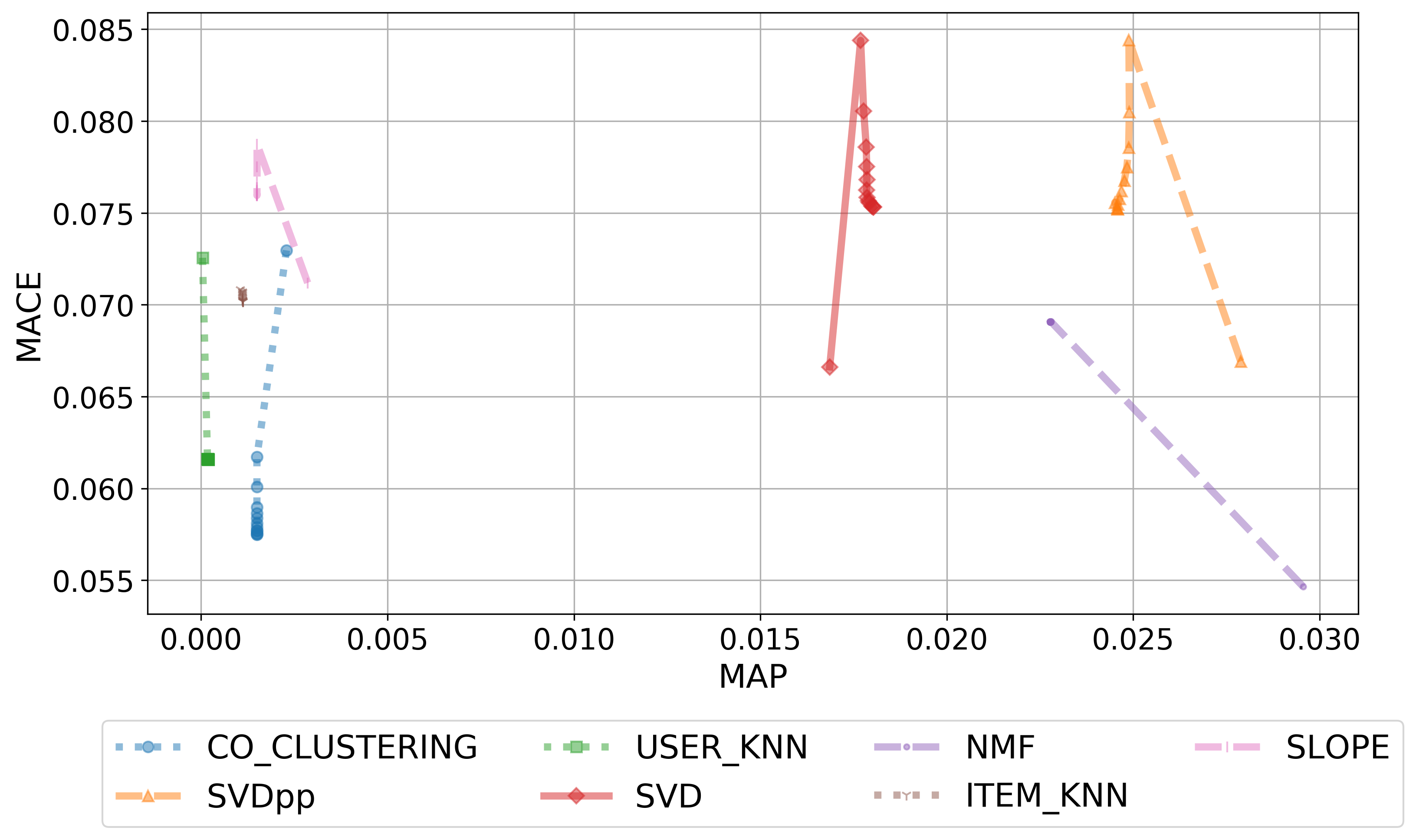}
		\caption{$\chi^2$ and $LIN$}
		\label{fig:ml_map_mace_chi_lin}
	\end{subfigure}
	\begin{subfigure}{0.48\textwidth}
		\centering
		\includegraphics[width=\linewidth]{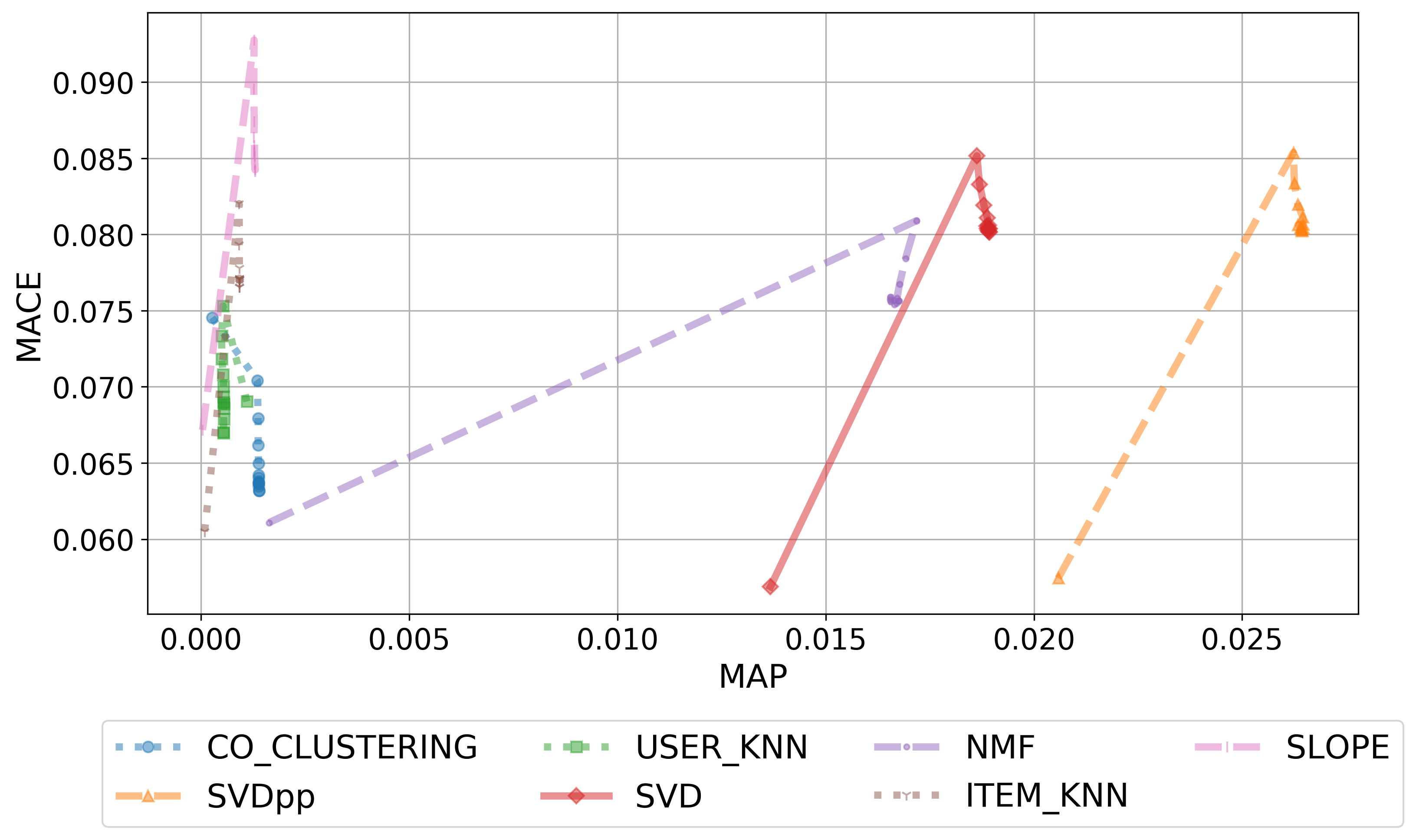}
		\caption{$\chi^2$ and $LOG$}
		\label{fig:ml_map_mace_chi_log}
	\end{subfigure}
	
	\caption{Movielens - MAP (x) and MACE (y) - Results from $LIN$ and $LOG$ trade-offs combined with the KL-divergence, Hellinger and Pearson Chi Square.}
	\label{fig:ml_map_mace}
\end{figure*}

We start the dataset analysis with the system combinations that implement the \textbf{KL-divergence}. Figures \ref{fig:ml_map_mace_kl_lin} and \ref{fig:ml_map_mace_kl_log} and Table \ref{tab:CCE_ML_MACE_MAP} (Lines 1 and 2) show that the best trade-off balance is the $LOG$. KL-LOG-SVD++ combination achieves the best performance. This find can be seen in Figure \ref{fig:ml_map_mace_kl_log}. The second and third places are the other matrix factorizations. As to the $LOG$ trade-off (Figure \ref{fig:ml_map_mace_kl_log}), the second and third places are achieved sequentially by SVD and NMF and to $LIN$ trade-off (Figure \ref{fig:ml_map_mace_kl_lin}) the recommenders change the positions to NMF and SVD sequentially. As indicated by the Table \ref{tab:CCE_ML_MACE_MAP} (Lines 1 and 2) the best performances, in the KL-divergence, are all matrix factorization approaches, specially to the SVD++ with $CCE=3.05$ ($LOG$) and $CCE=3.52$ ($LIN$). The worst results are held by the KNNs approaches that consider the KL-divergence as a fixed divergence measure.

The results achieved by the \textbf{Hellinger} combinations are shown in Figure \ref{fig:ml_map_mace_he_lin}, in Figure \ref{fig:ml_map_mace_he_log} and in Table \ref{tab:CCE_ML_MACE_MAP} (Lines 3 and 4). The results indicate that the $LOG$ trade-off (Figure \ref{fig:ml_map_mace_he_log}), similar to the KL analysis, achieves the best performance and such behavior remain in all matrix factorization approaches. It is possible to verify that the worst performance remains with the KNNs approaches. The Hellinger achieves lower MACE values than the KL-divergence, indicating that the HE produces a recommended list with a low degree of calibration error. The results on Table \ref{tab:CCE_ML_MACE_MAP} (Lines 3 and 4) show that the SVD++, in the Hellinger, achieved the $CCE=2.08$ to $LOG$ and $CCE=2.38$ to $LIN$.

The last MAPxMACE results that we will analyze are the \textbf{Pearson Chi Square} ($\chi^2$) combinations. Figures \ref{fig:ml_map_mace_chi_lin} and \ref{fig:ml_map_mace_chi_log} with Table \ref{tab:CCE_ML_MACE_MAP} (Lines 5 and 6) show that the $\chi^2$ follows the same behavior found in the two previous measures analyzed. The best performances are all matrix factorization approaches, as the KNNs obtain the worst performance. The best combination performance, in the $\chi^2$ is obtained by the CHI-LIN-NMF with a $CCE=2.92$ (Table \ref{tab:CCE_ML_MACE_MAP} Lines 5 and 6), followed by the SVD++ in second place and the SVD in third place. In the $LOG$ trade-off the places follow the sequence SVD++, SVD and NMF.

The results indicate that if the system is focused on MAP and MACE and according to our experimental setup, that the best combination to be implemented in the Movielens dataset is the HE-LOG-SVD++ with a $CCE=2.08$ as presented in Table \ref{tab:CCE_ML_MACE_MAP} Line 4. We note that the $LOG$ trade-off is a part of the proposal of this research. The second best performance based on the CCE belongs to the HE-LIN-SVD++ with $CCE=2.38$ and the third place belongs to HE-LIN-NMF with $CCE=2.51$ (Table \ref{tab:CCE_ML_MACE_MAP} Line 3). We also can observe that the Hellinger divergence measure achieves all best performances on the CCE. We can also observe that there is no correlation between the MAP and MACE.

\subsubsection{Taste Profile}
\begin{table*}[t!]
	\centering
	
	\caption{Taste Profile - CCE - Results from $LIN$ and $LOG$ trade-offs combined with the KL, HE and CHI divergence measure for each recommender algorithm.}
	\label{tab:CCE_OMS_MACE_MAP}
	\resizebox{\textwidth}{!}{%
    \begin{tabular}{c|c|ccccccc}
    \hline
    \textbf{Divergence} & \textbf{\textbf{KL}} & \textbf{\textbf{Co Clustering}} & \textbf{\textbf{Item KNN}} & \textbf{\textbf{NMF}} & \textbf{\textbf{Slope One}} & \textbf{\textbf{SVD}} & \textbf{\textbf{SVD++}} & \textbf{\textbf{User KNN}} \\ \hline
    KL                  & $LIN$                & 5093.61                         & {\ul \textbf{25.73}}       & 667.67                & 1471.03                     & 489.45                & 329.39                  & 1175.62                    \\
    KL                  & $LOG$                & 5199.86                         & {\ul \textbf{26.68}}       & 2307.22               & 1714.39                     & 721.64                & 559.37                  & 1232.02                    \\ \hline
    HE                  & $LIN$                & 5072.75                         & {\ul \textbf{14.39}}       & 459.35                & 1406.45                     & 303.95                & 278.98                  & 1148.17                    \\
    HE                  & $LOG$                & 5057.98                         & {\ul \textbf{21.32}}       & 2336.09               & 1709.8                      & 709.28                & 552.41                  & 1234.66                    \\ \hline
    CHI                 & $LIN$                & 4308.4                          & {\ul \textbf{23.29}}       & 469.9                 & 1438.43                     & 308.4                 & 283.38                  & 1047.84                    \\
    CHI                 & $LOG$                & 4442.51                         & {\ul \textbf{27.64}}       & 2287.83               & 1712.22                     & 659.49                & 531.97                  & 1223.63                    \\ \hline
    \end{tabular}%
    }
\end{table*}

\begin{figure*}[ht!]
	\centering
	\begin{subfigure}{0.48\textwidth}
		\centering
		\includegraphics[width=\linewidth]{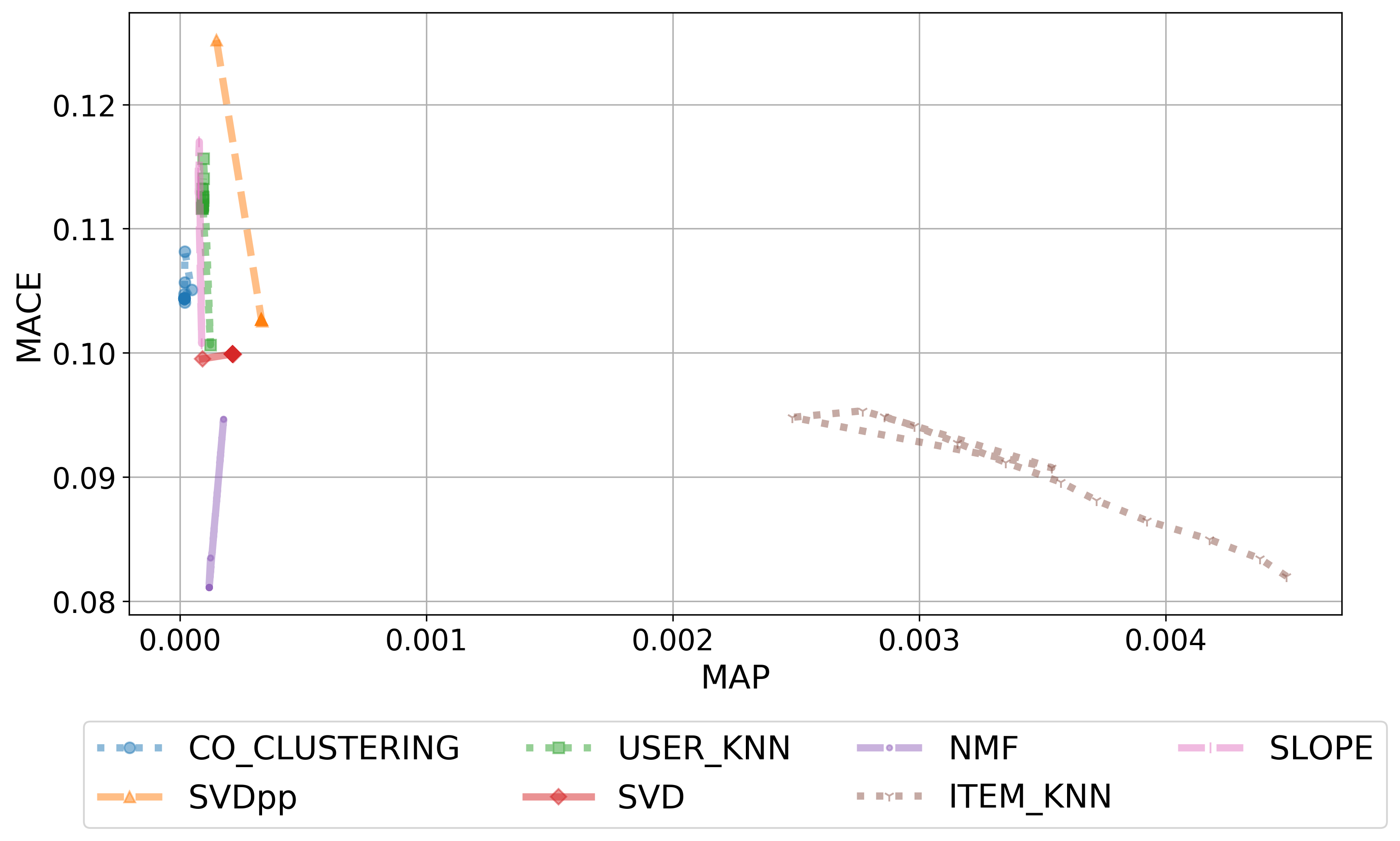}
		\caption{KL and $LIN$}
		\label{fig:oms_map_mace_kl_lin}
	\end{subfigure}
	\begin{subfigure}{0.48\textwidth}
		\centering
		\includegraphics[width=\linewidth]{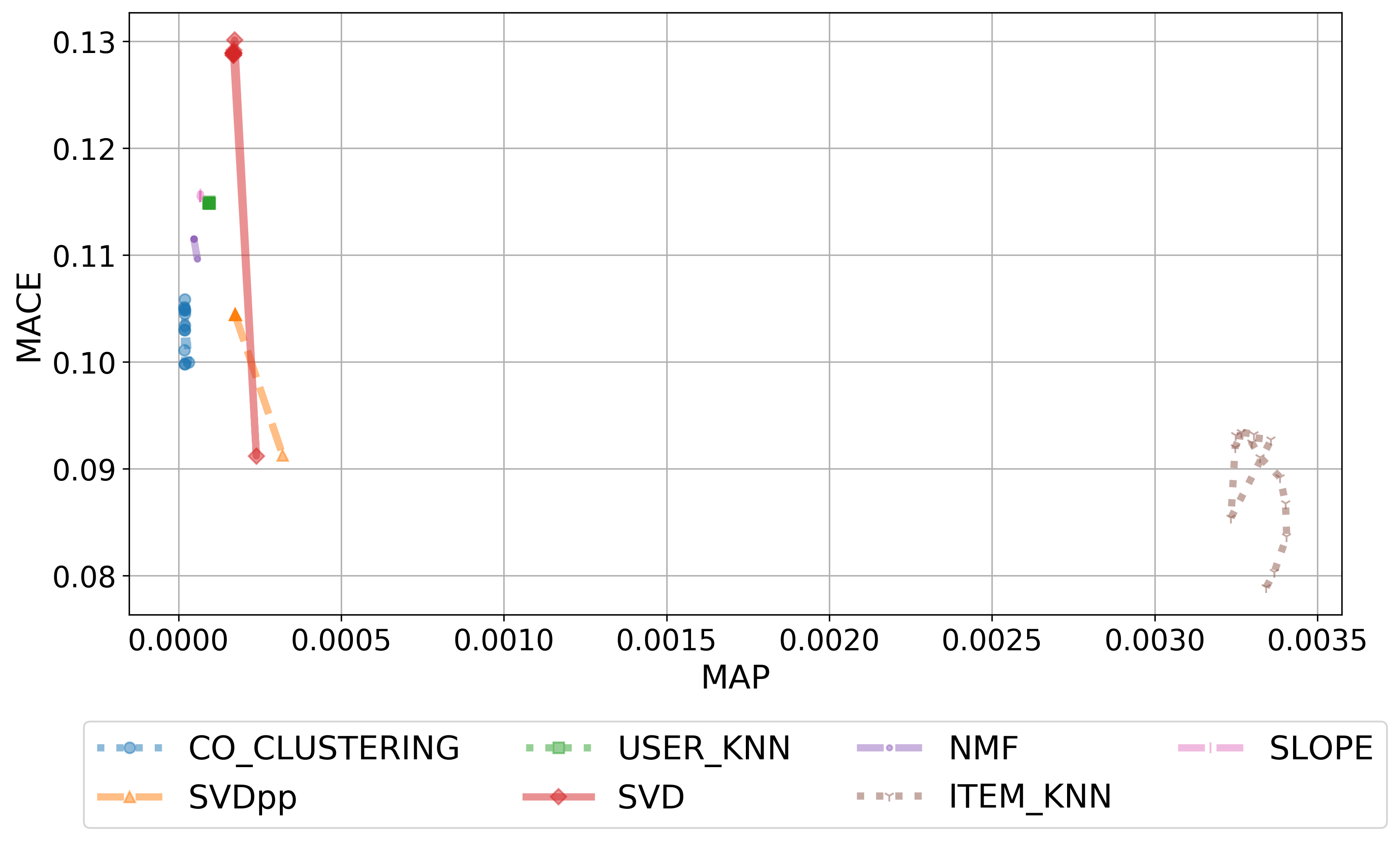}
		\caption{KL and $LOG$}
		\label{fig:oms_map_mace_kl_log}
	\end{subfigure}
	~

	\begin{subfigure}{0.48\textwidth}
		\centering
		\includegraphics[width=\linewidth]{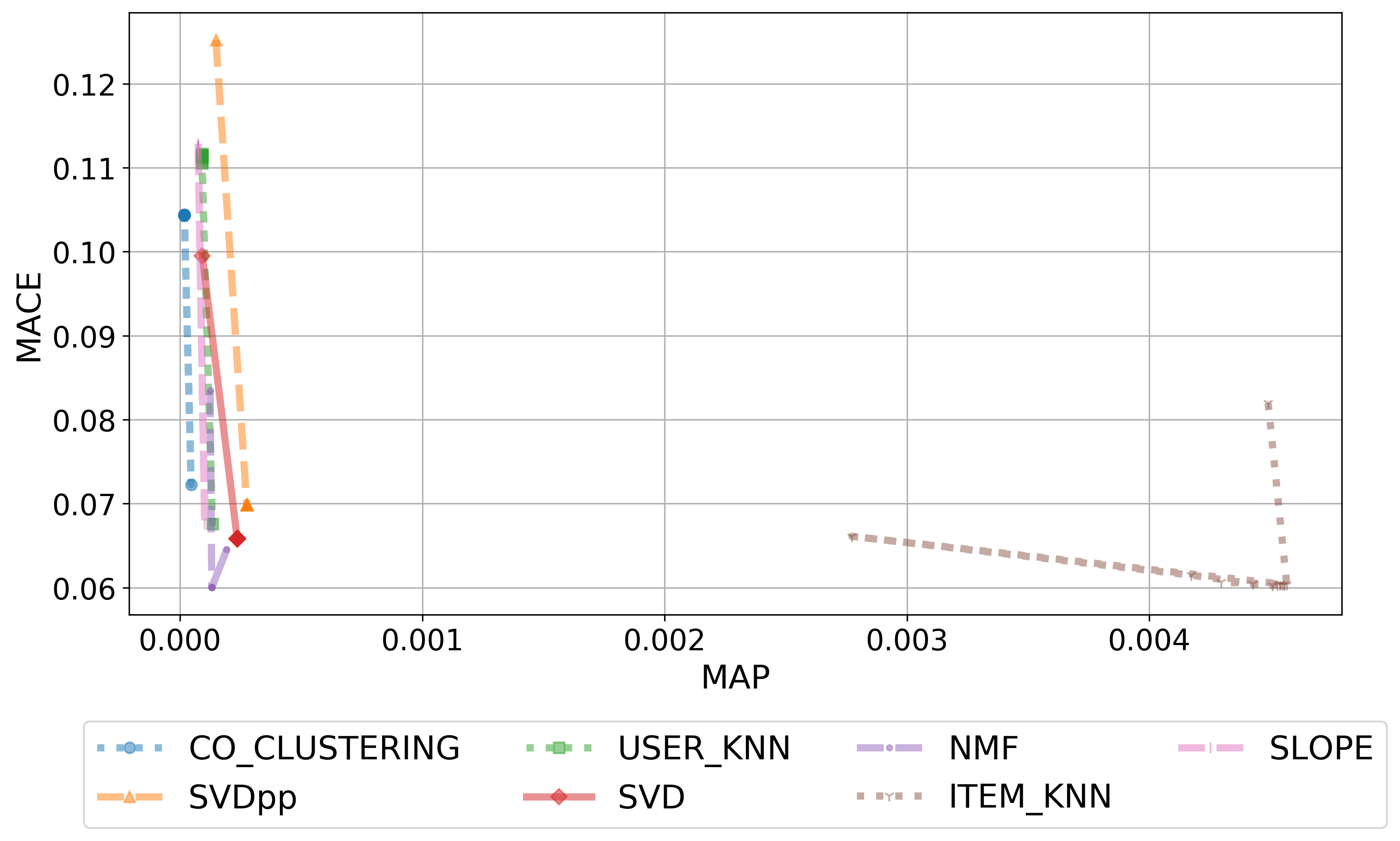}
		\caption{Hellinger and $LIN$}
		\label{fig:oms_map_mace_he_lin}
	\end{subfigure}
	\begin{subfigure}{0.48\textwidth}
		\centering
		\includegraphics[width=\linewidth]{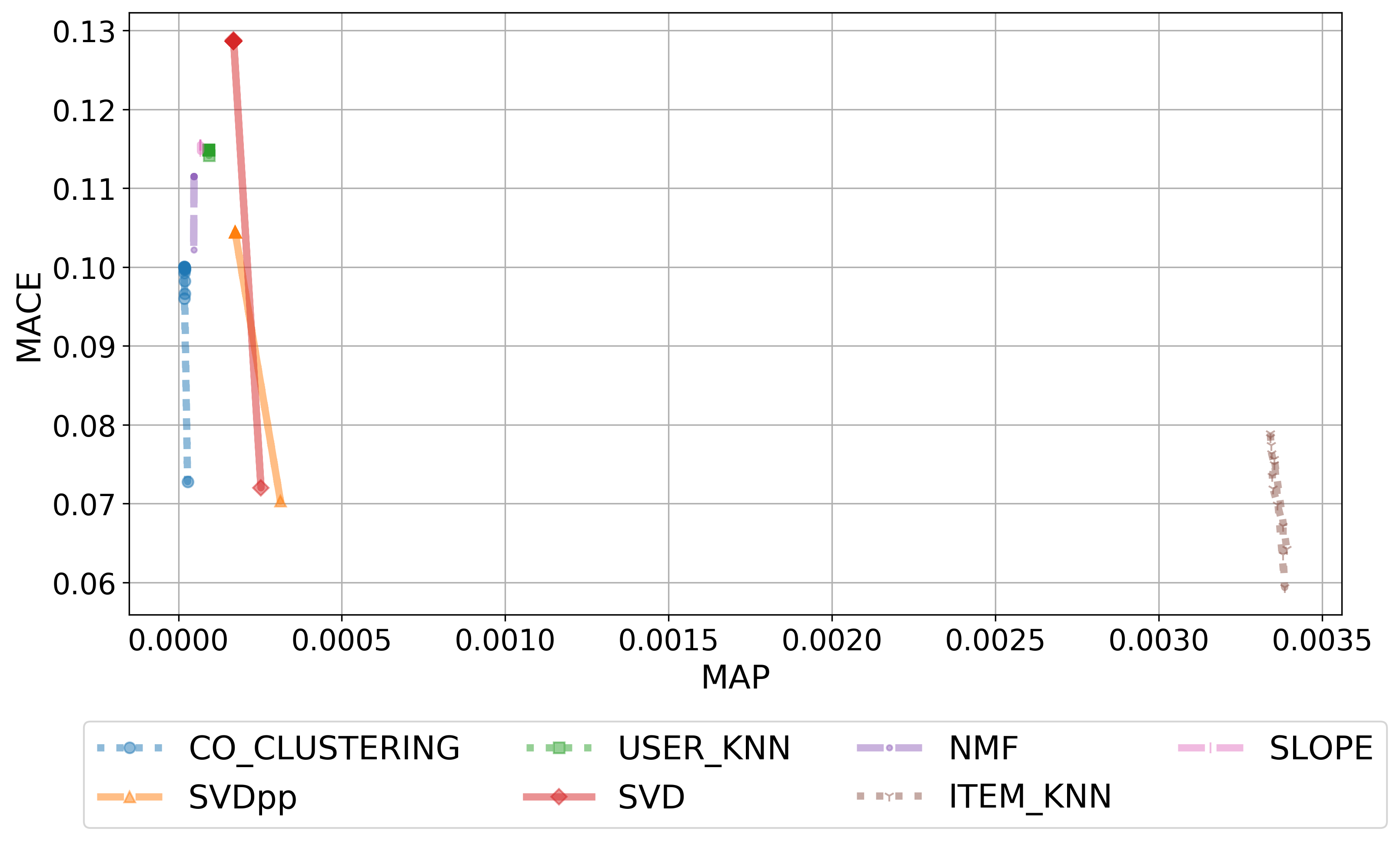}
		\caption{Hellinger and $LOG$}
		\label{fig:oms_map_mace_he_log}
	\end{subfigure}
	
	~
	
	\begin{subfigure}{0.48\textwidth}
		\centering
		\includegraphics[width=\linewidth]{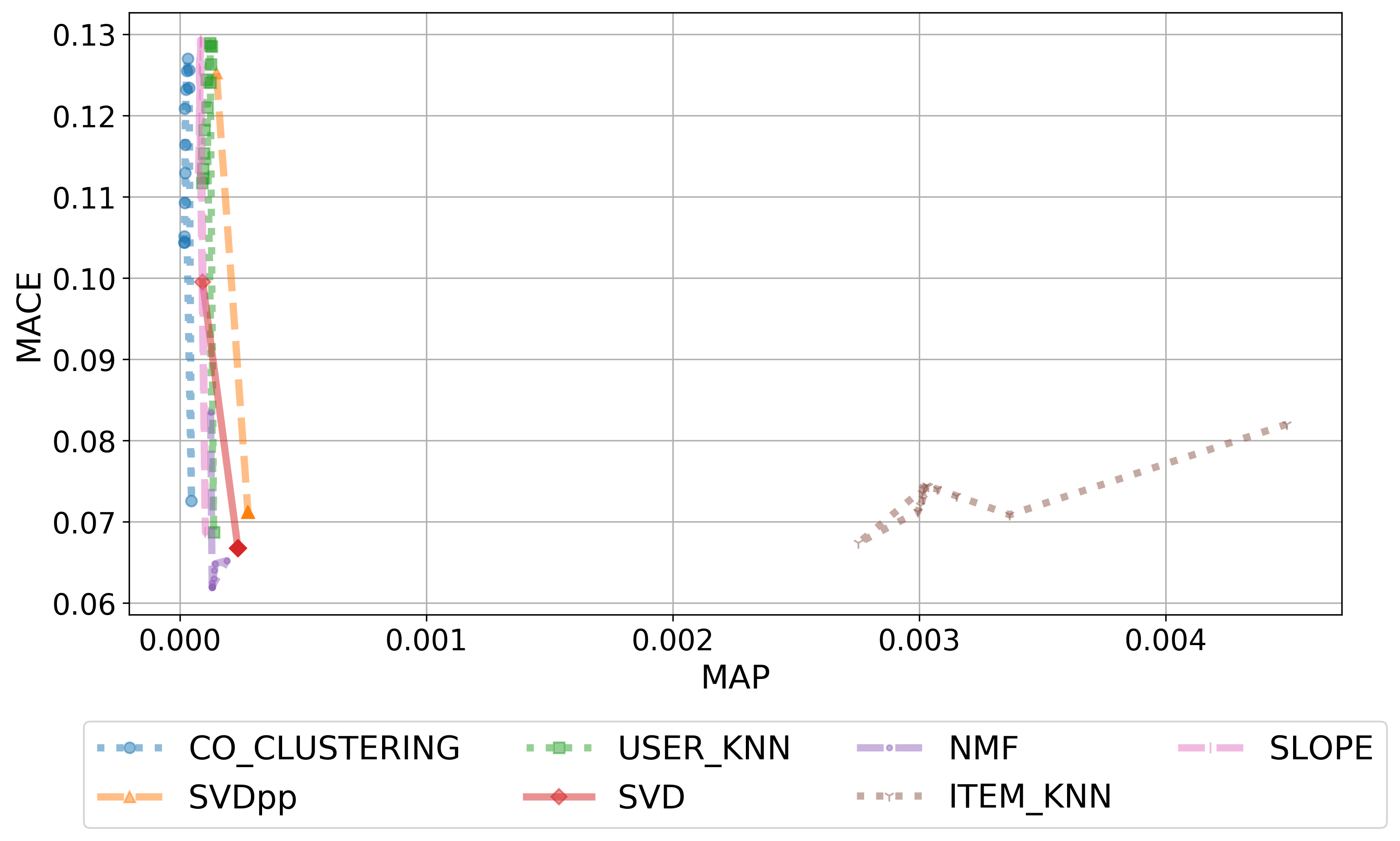}
		\caption{$\chi^2$ and $LIN$}
		\label{fig:oms_map_mace_chi_lin}
	\end{subfigure}
	\begin{subfigure}{0.48\textwidth}
		\centering
		\includegraphics[width=\linewidth]{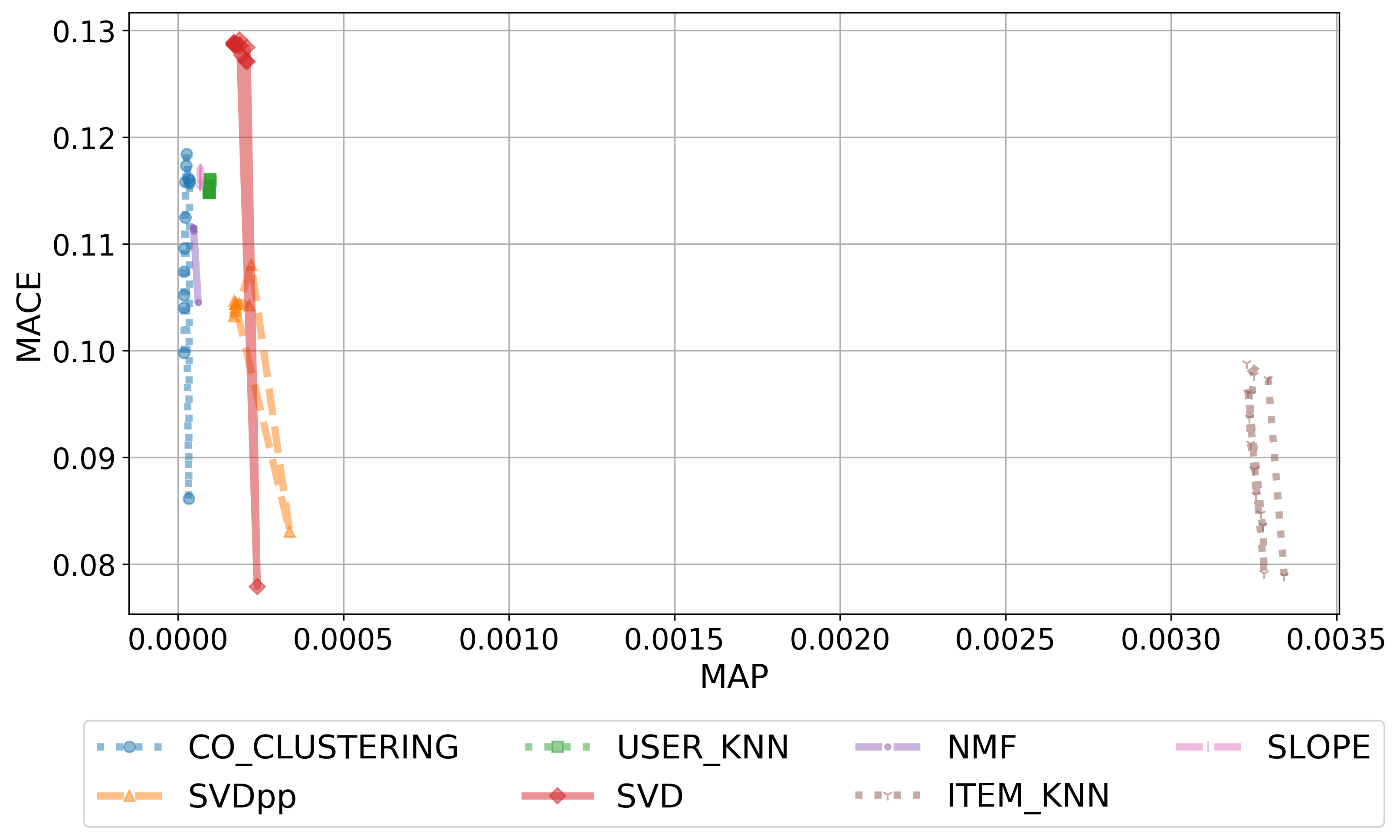}
		\caption{$\chi^2$ and $LOG$}
		\label{fig:oms_map_mace_chi_log}
	\end{subfigure}
	
	\caption{Taste Profile - MAP (x) and MACE (y) - Results from $LIN$ and $LOG$ trade-offs combined with the KL-divergence, Hellinger and Pearson Chi Square.}
	\label{fig:oms_map_mace}
\end{figure*}

We will start our results analysis in the Taste Profile by the \textbf{KL-divergence}. Figures \ref{fig:oms_map_mace_kl_lin} and \ref{fig:oms_map_mace_kl_log} and Table \ref{tab:CCE_OMS_MACE_MAP} (Lines 1 and 2) show that the best recommender performance, in the KL-divergence, belongs to the Item-KNN both in $LIN$ and $LOG$. It is possible to verify that the Item-KNN results in the $LIN$ have more variability than in the $LOG$ trade-off. The CCE obtained by the combination KL-LIN-Item-KNN is $25.73$ against $26.68$ from KL-LOG-Item-KNN (Table \ref{tab:CCE_OMS_MACE_MAP} Lines 1 and 2). The second and third place are the SVD++ and SVD, both matrix factorization approaches. The worst performance belongs to Co Clustering.

Figures \ref{fig:oms_map_mace_he_lin} and \ref{fig:oms_map_mace_he_log} support our analysis of the divergence measure \textbf{Hellinger} in the Taste Profile, besides Table \ref{tab:CCE_OMS_MACE_MAP} in Lines 3 and 4 presents the CCE performances. Similar to KL-divergence on Taste Profile, the Item-KNN obtains the best performances in the $LIN$ with $CCE=14.39$ and $LOG$ with $CCE=21.32$, as well as the $LOG$ trade-off has less variability in the results when compared with the $LIN$. It is possible to verify, through Table \ref{tab:CCE_OMS_MACE_MAP} Lines 3 and 4, that the second and third place are the SVD++ and SVD, both matrix factorization approaches. The worst performance remains with the Co Clustering approach.

The last results to the Taste Profile are the \textbf{Pearson Chi Square} ($\chi^2$) combinations. Figures \ref{fig:oms_map_mace_chi_lin} and \ref{fig:oms_map_mace_chi_log} combined with Table \ref{tab:CCE_OMS_MACE_MAP} Lines 5 and 6 present the results. The $\chi^2$ follows the same behavior found in the two previous measures analyzed in the dataset. The best performance belongs to Item-KNN. The $LOG$ trade-off has less variability in the performance and the $LIN$ obtained the best performance in the trade-off balances. The CHI-LIN-Item-KNN obtained the $CCE=23.29$ and the CHI-LOG-Item-KNN obtained $CCE=27.64$. The SVD++ obtained the second place and the SVD the third place in both trade-off balances.

From the CCE results in the Taste Profile, we can observe that if the system is focused on MAP and MACE, the best performance refers to the combination HE-LIN-Item-KNN with a $CCE=14.39$, followed by the combination HE-LOG-Item-KNN with a $CCE=21.32$. These results can be verified in Table \ref{tab:CCE_OMS_MACE_MAP} at lines 3 and 4. We remind that results are restricted to the CCE (MAP and MACE metric crossed).

\begin{table*}[!t]
	\centering

	\caption{Movielens - CMC - Results from $LIN$ and $LOG$ trade-offs combined with the KL, HE and CHI divergence measure for each recommender algorithm.}
	\label{tab:CMC_ML_MRMC_MAP}
	\resizebox{\textwidth}{!}{%
    \begin{tabular}{c|c|ccccccc}
        \hline
        \textbf{Divergence} & \textbf{Trade-off} & \textbf{Co Clustering} & \textbf{Item KNN} & \textbf{NMF} & \textbf{Slope One} & \textbf{SVD} & \textbf{SVD++} & \textbf{User KNN} \\ \hline
        KL                  & $LIN$              & 60.44                           & 90.73                      & 3.8                   & 63.52                       & 4.95                  & {\ul \textbf{3.52}}     & 430.91                     \\
        KL                  & $LOG$              & 62.94                           & 110.46                     & 5.76                  & 83.46                       & 4.31                  & {\ul \textbf{3.05}}     & 130.58                     \\ \hline
        HE                  & $LIN$              & 298.94                          & 454.46                     & {\ul \textbf{20.22}}  & 278.07                      & 31.78                 & 20.56                   & 5319.06                    \\
        HE                  & $LOG$              & 576.33                          & 1654.34                    & 83.42                 & 982.12                      & 31.01                 & {\ul \textbf{20.95}}    & 616.98                     \\ \hline
        CHI                 & $LIN$              & 156.53                          & 228.18                     & 10.52                 & 136.21                      & 13.21                 & {\ul \textbf{9.39}}     & 1367.03                    \\
        CHI                 & $LOG$              & 184.39                          & 294.33                     & 15.84                 & 181.31                      & 12.8                  & {\ul \textbf{9.08}}     & 391.45                     \\ \hline
    \end{tabular}%
    }
\end{table*}

\begin{figure*}[!ht]
	\centering
	\begin{subfigure}{0.48\textwidth}
		\centering
		\includegraphics[width=\linewidth]{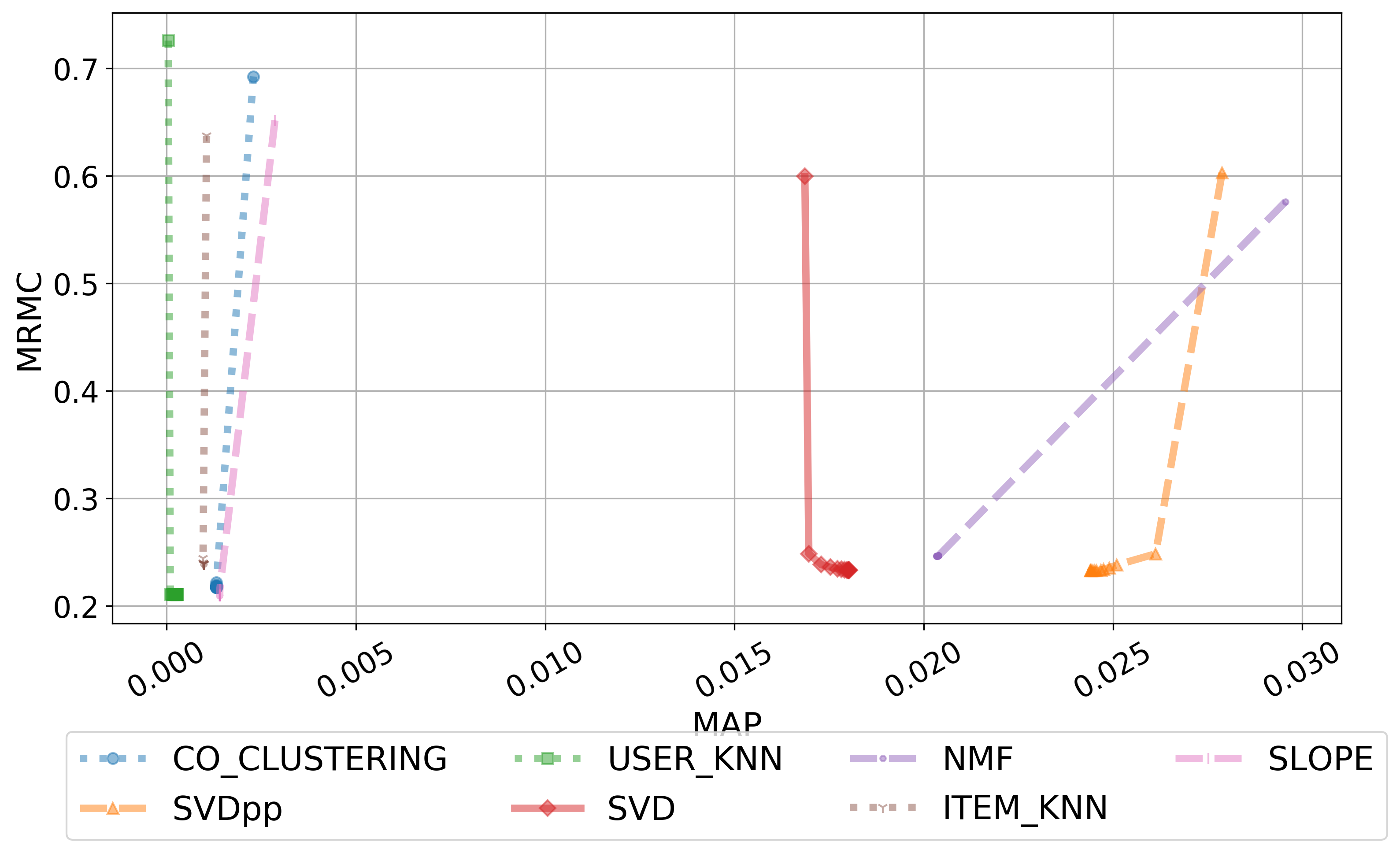}
		\caption{KL and $LIN$}
		\label{fig:ml_map_mrmc_kl_lin}
	\end{subfigure}	
	\begin{subfigure}{0.48\textwidth}
		\centering
		\includegraphics[width=\linewidth]{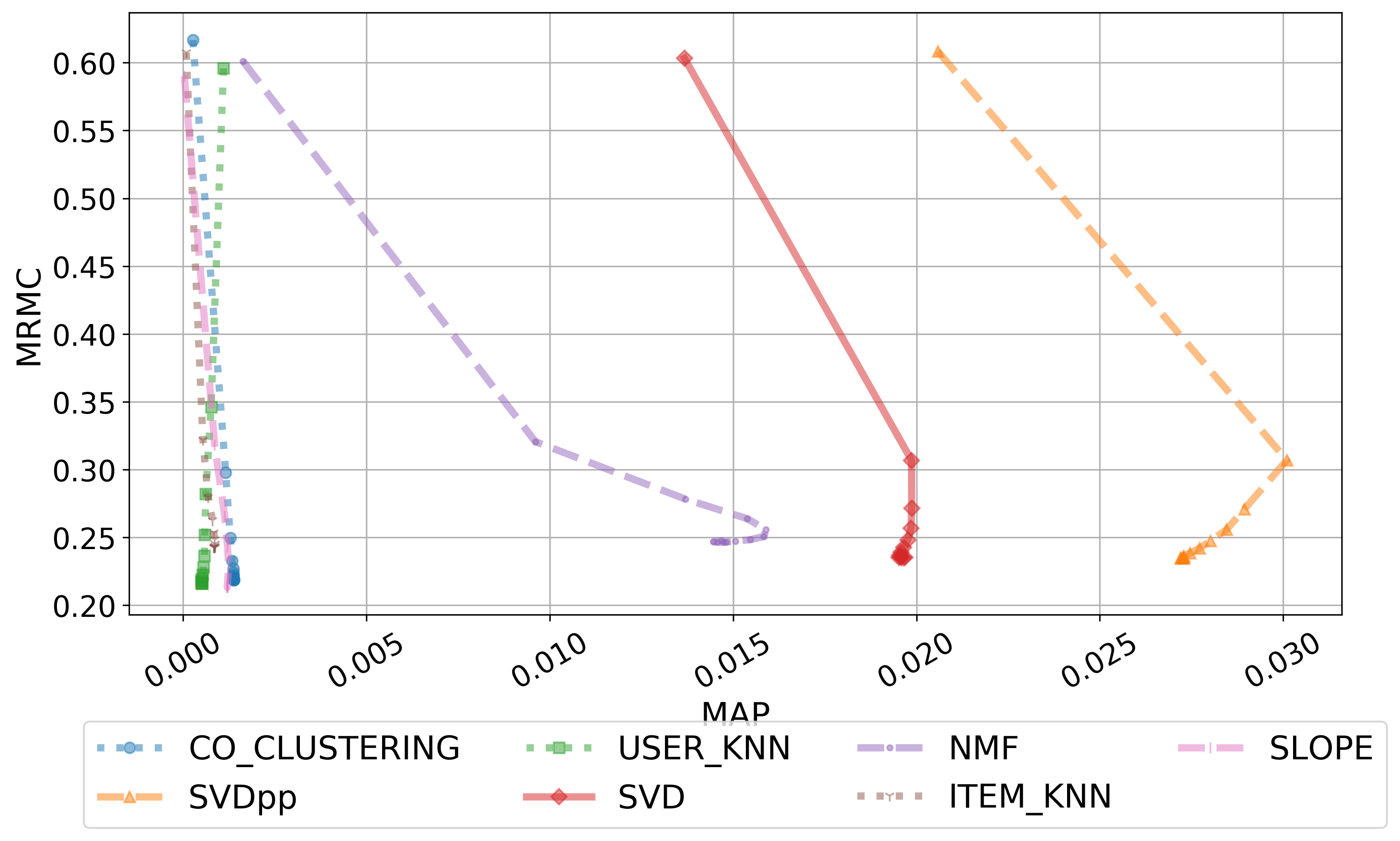}
		\caption{KL and $LOG$}
		\label{fig:ml_map_mrmc_kl_log}
	\end{subfigure}
	
	~

	\begin{subfigure}{0.48\textwidth}
		\centering
		\includegraphics[width=\linewidth]{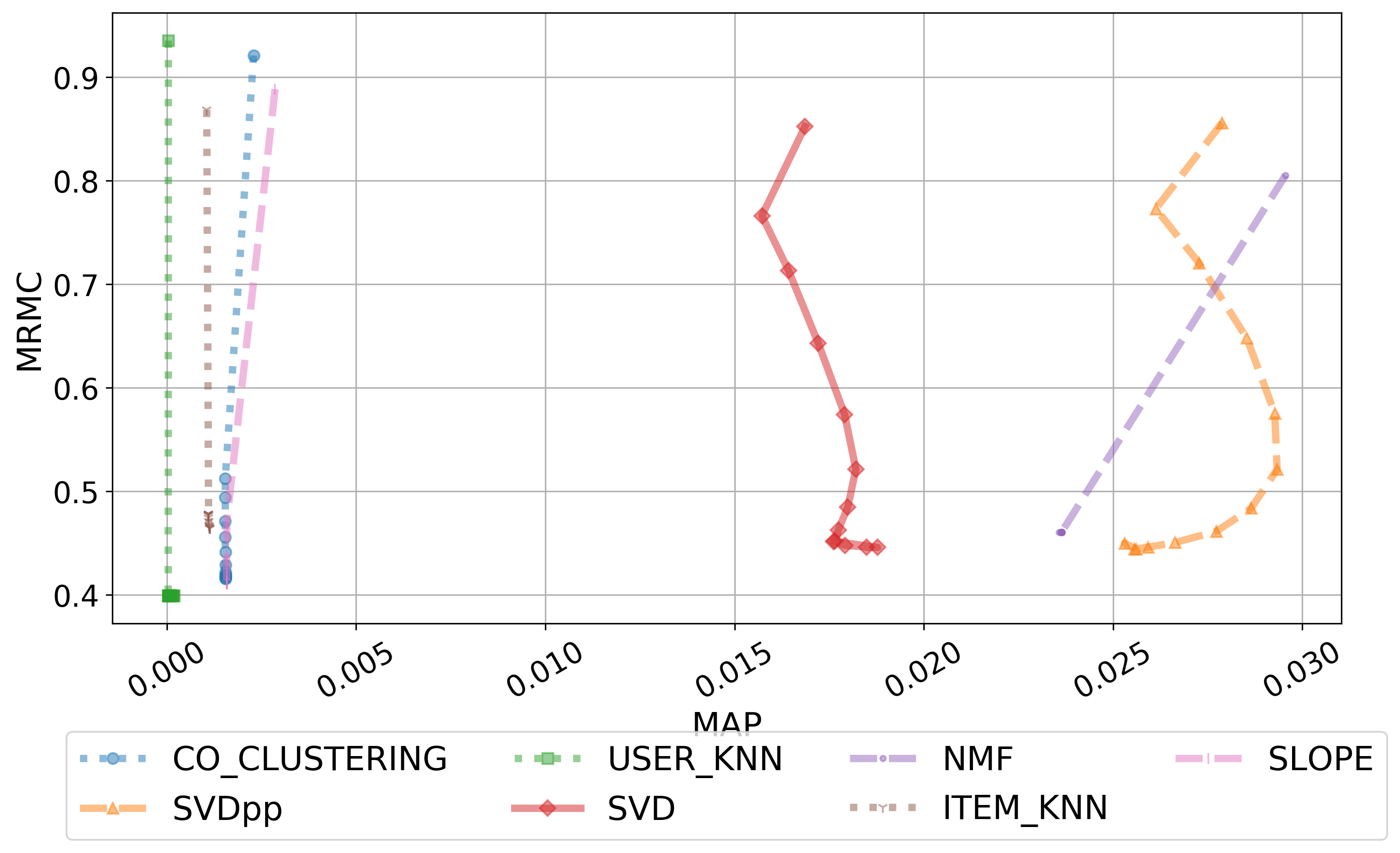}
		\caption{Hellinger and $LIN$}
		\label{fig:ml_map_mrmc_he_lin}
	\end{subfigure}
	\begin{subfigure}{0.48\textwidth}
		\centering
		\includegraphics[width=\linewidth]{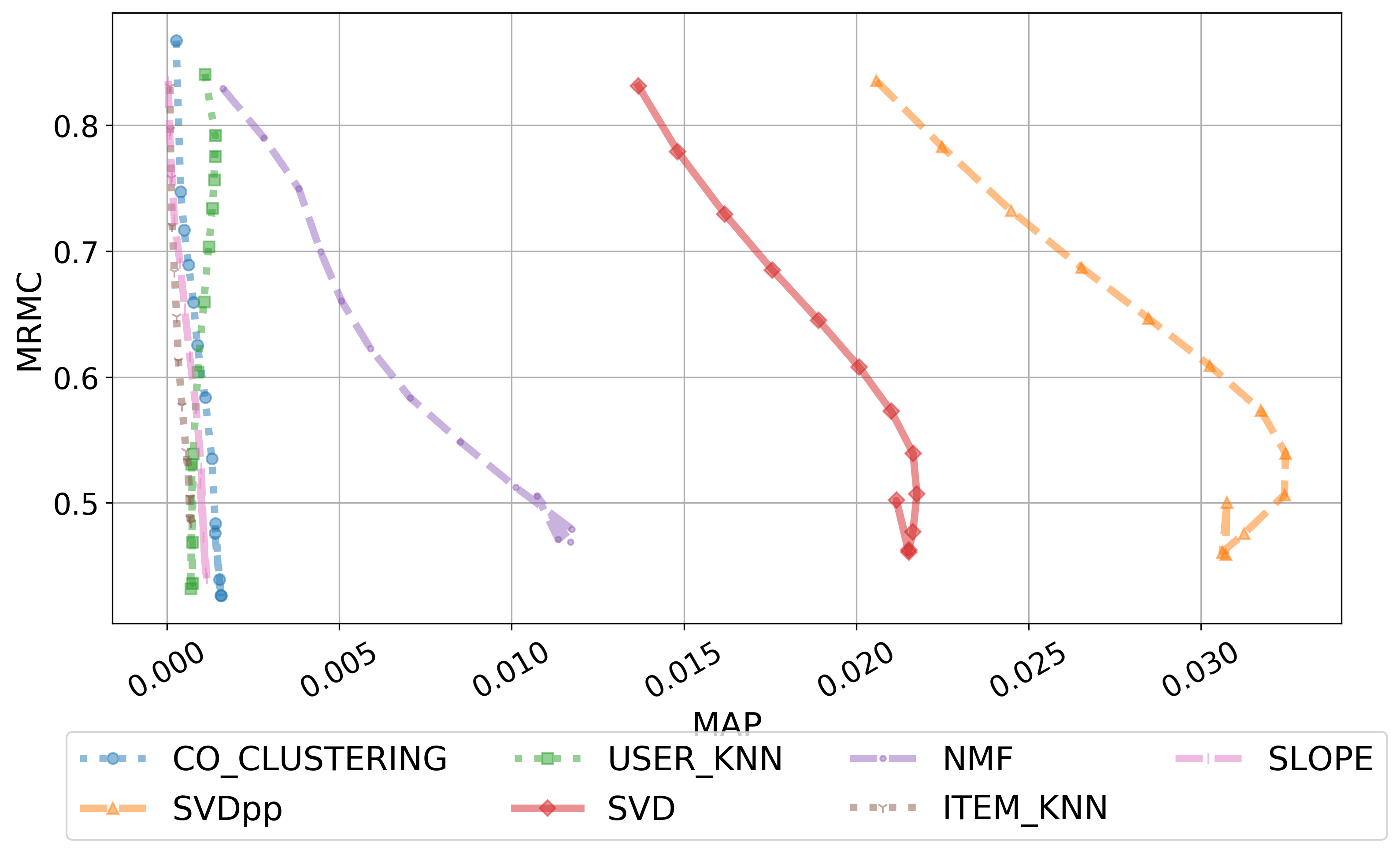}
		\caption{Hellinger and $LOG$}
		\label{fig:ml_map_mrmc_he_log}
	\end{subfigure}
	
	~
	
	\begin{subfigure}{0.48\textwidth}
		\centering
		\includegraphics[width=\linewidth]{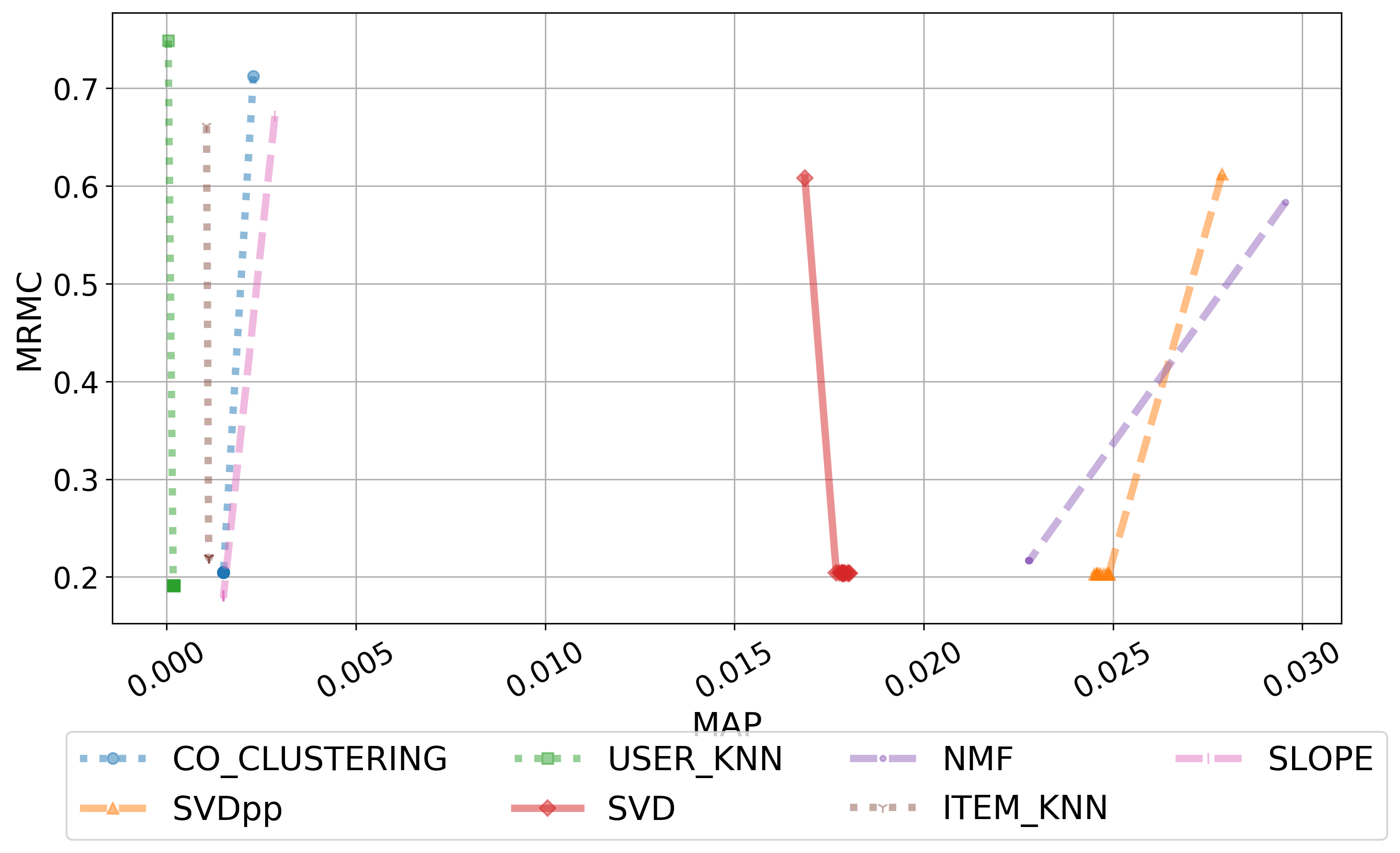}
		\caption{$\chi^2$ and $LIN$}
		\label{fig:ml_map_mrmc_chi_lin}
	\end{subfigure}
	\begin{subfigure}{0.48\textwidth}
		\centering
		\includegraphics[width=\linewidth]{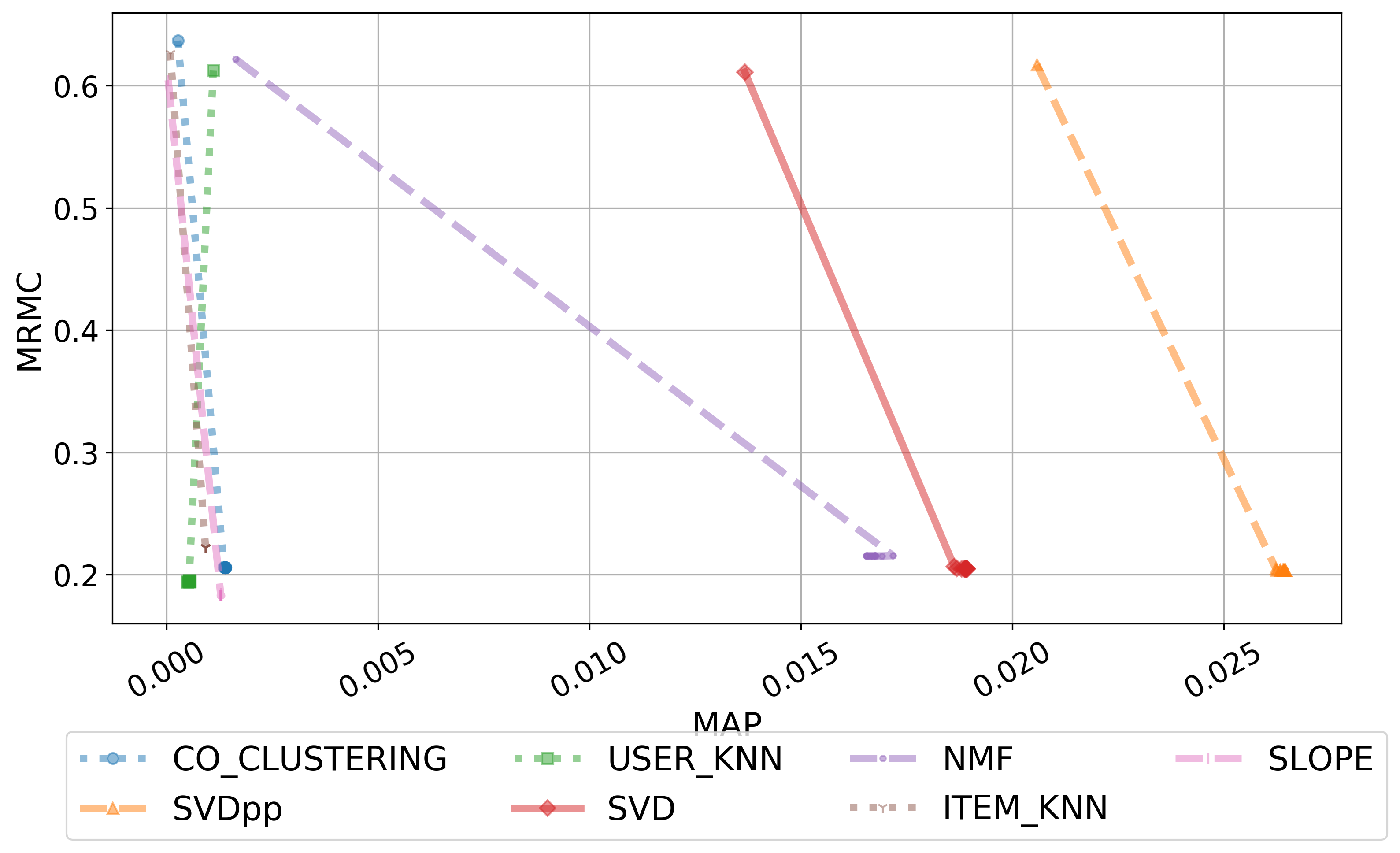}
		\caption{$\chi^2$ and $LOG$}
		\label{fig:ml_map_mrmc_chi_log}
	\end{subfigure}
	
	\caption{Movielens - MAP (x) and MRMC (y) - Results from $LIN$ and $LOG$ trade-offs combined with the KL-divergence, Hellinger and Pearson Chi Square.}
	\label{fig:ml_map_mrmc}
\end{figure*}

\subsection{MAP x MRMC}
\label{sec:sub:map_mrmc}
\begin{figure*}[t!]
	\centering
	\begin{subfigure}{0.48\textwidth}
		\centering
		\includegraphics[width=\linewidth]{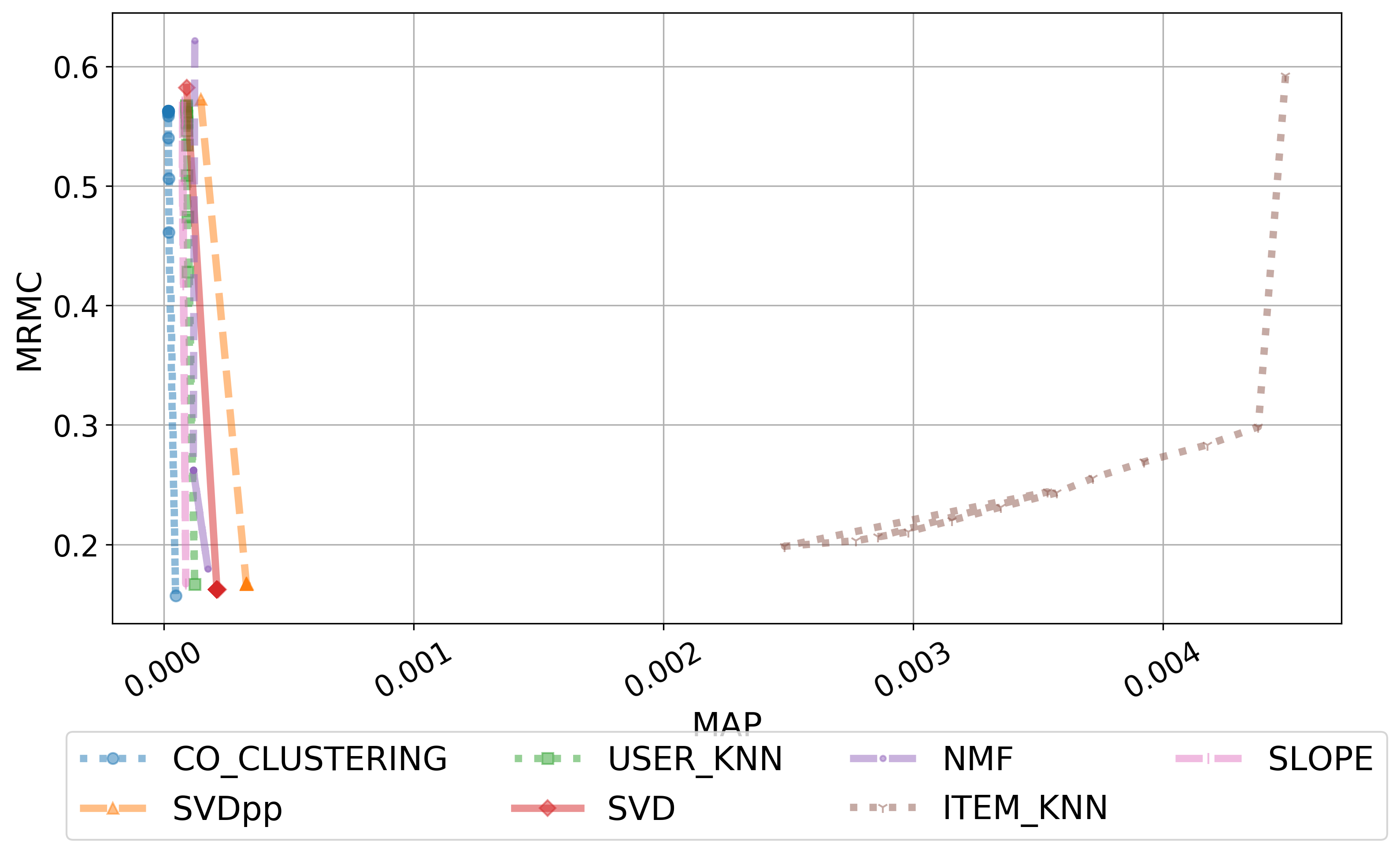}
		\caption{KL and $LIN$}
		\label{fig:oms_map_mrmc_kl_lin}
	\end{subfigure}
	\begin{subfigure}{0.48\textwidth}
		\centering
		\includegraphics[width=\linewidth]{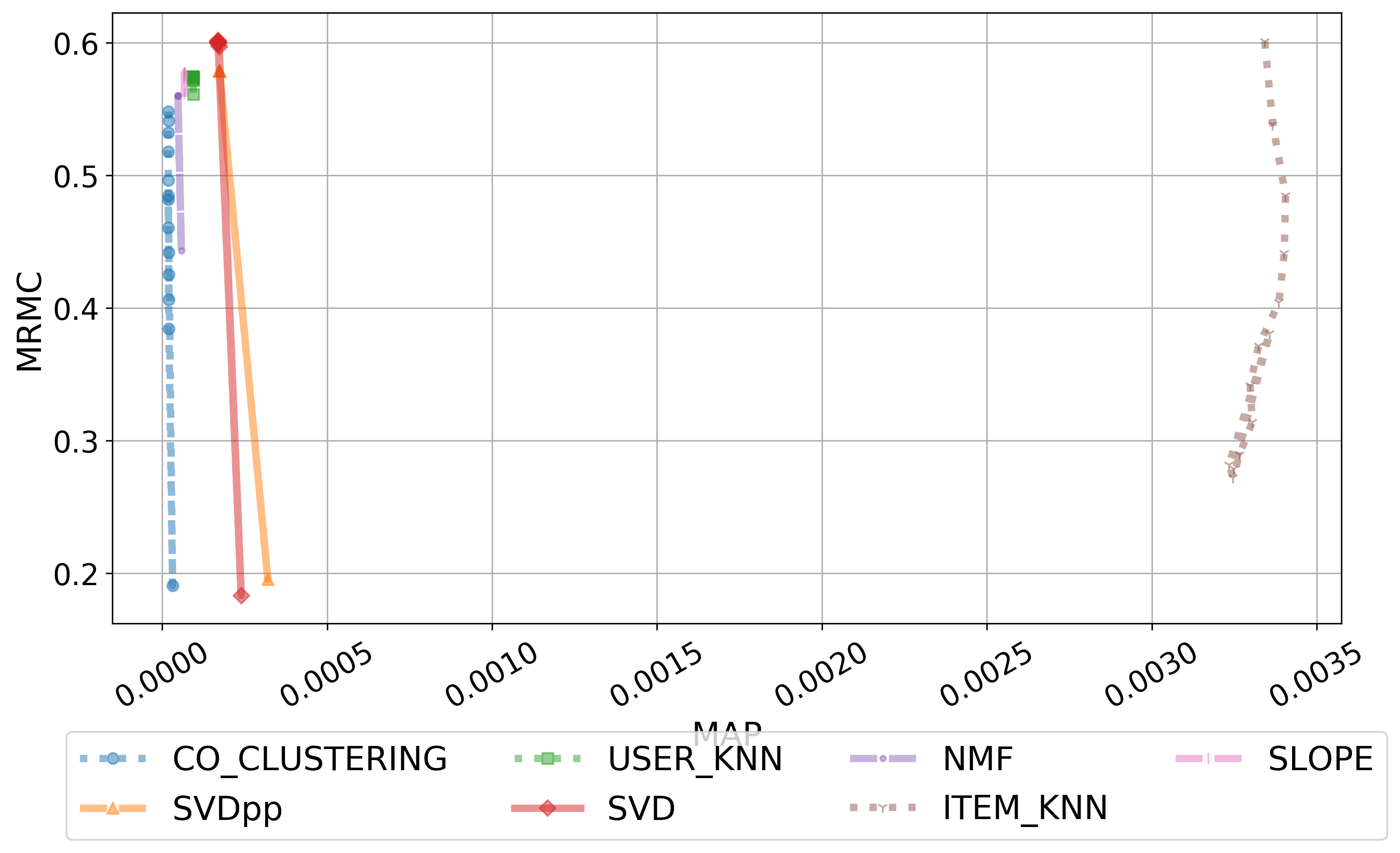}
		\caption{KL and $LOG$}
		\label{fig:oms_map_mrmc_kl_log}
	\end{subfigure}
	
	~
	
	\begin{subfigure}{0.48\textwidth}
		\centering
		\includegraphics[width=\linewidth]{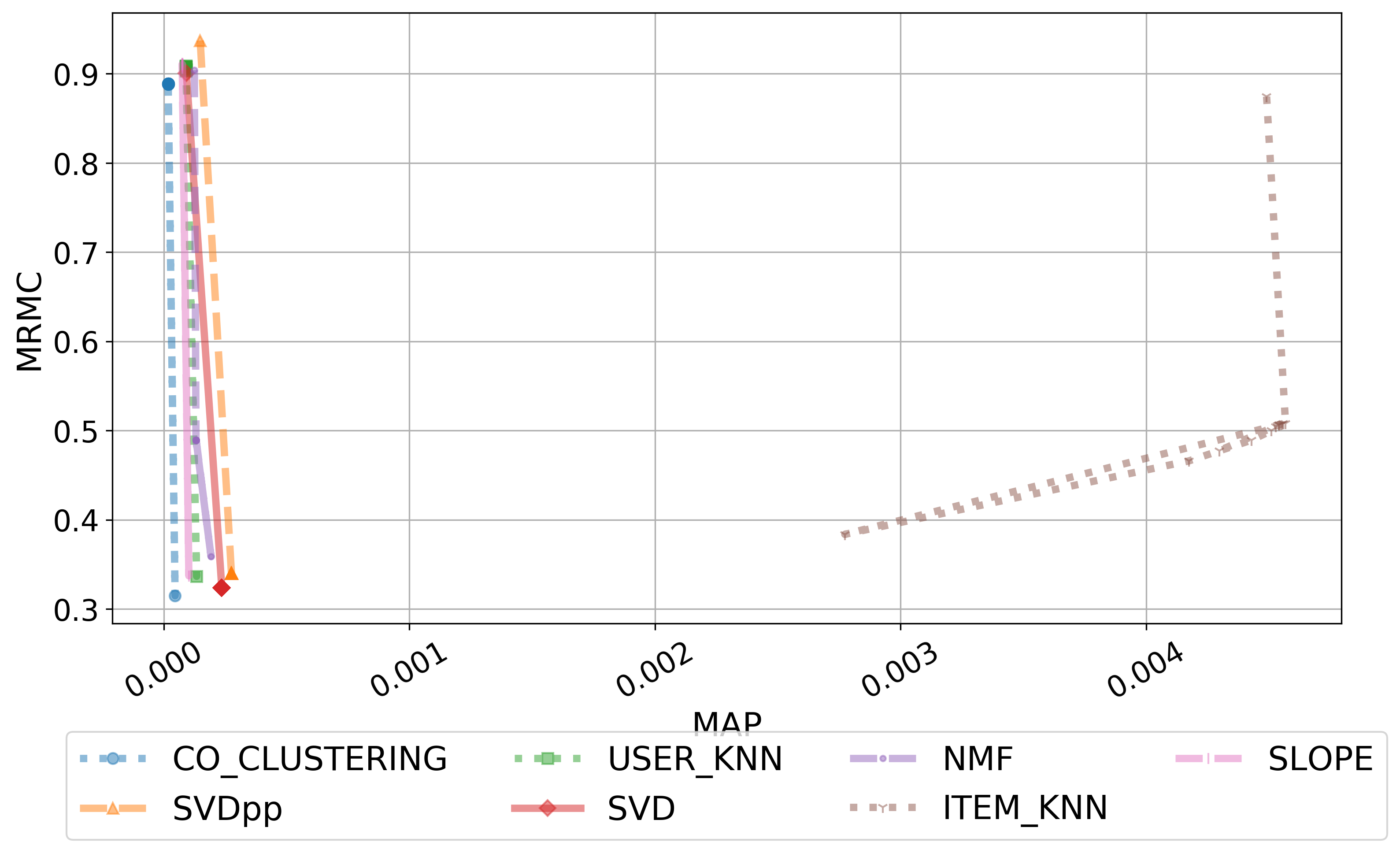}
		\caption{Hellinger and $LIN$}
		\label{fig:oms_map_mrmc_he_lin}
	\end{subfigure}
	\begin{subfigure}{0.48\textwidth}
		\centering
		\includegraphics[width=\linewidth]{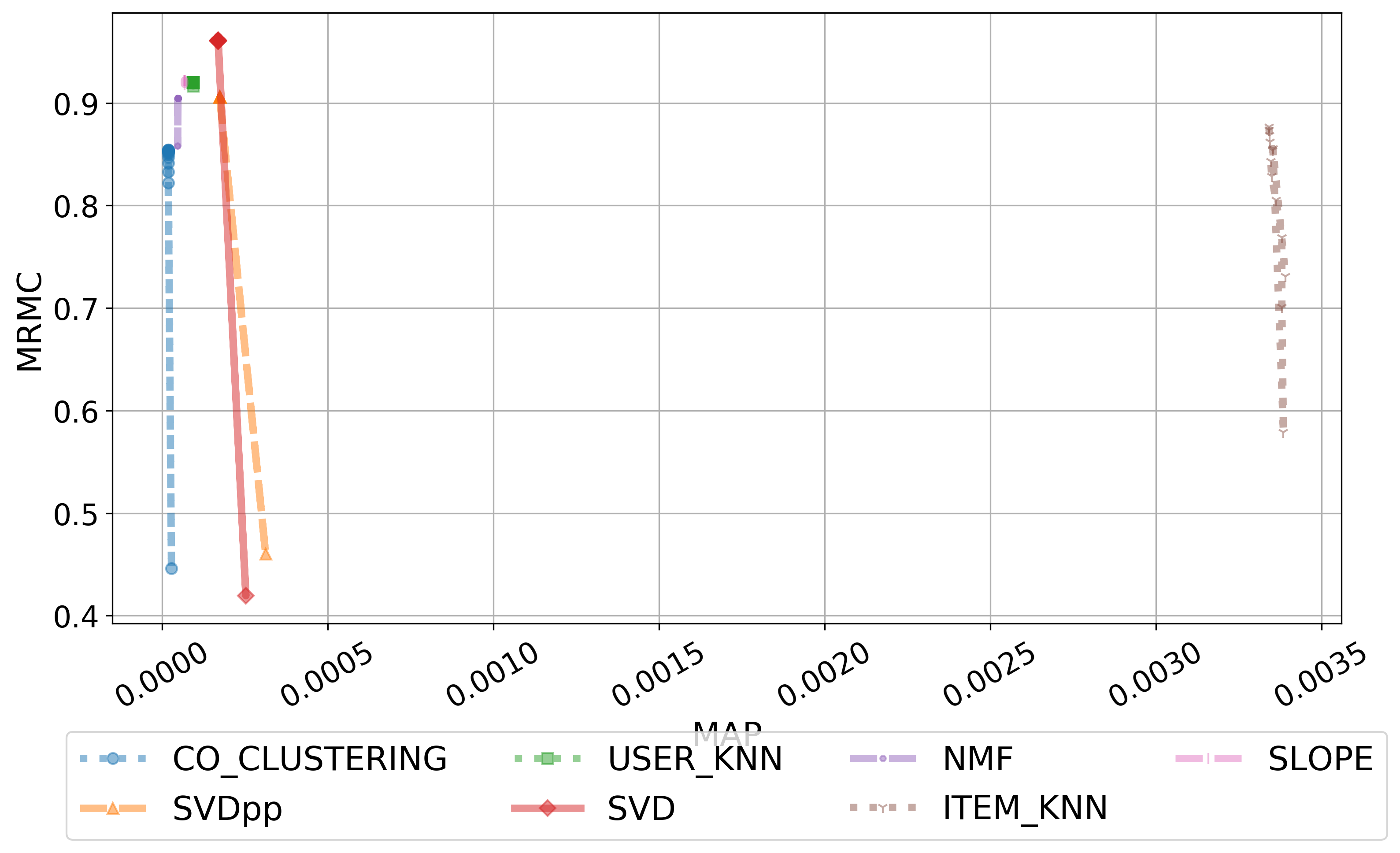}
		\caption{Hellinger and $LOG$}
		\label{fig:oms_map_mrmc_he_log}
	\end{subfigure}
	
	~
	
	\begin{subfigure}{0.48\textwidth}
		\centering
		\includegraphics[width=\linewidth]{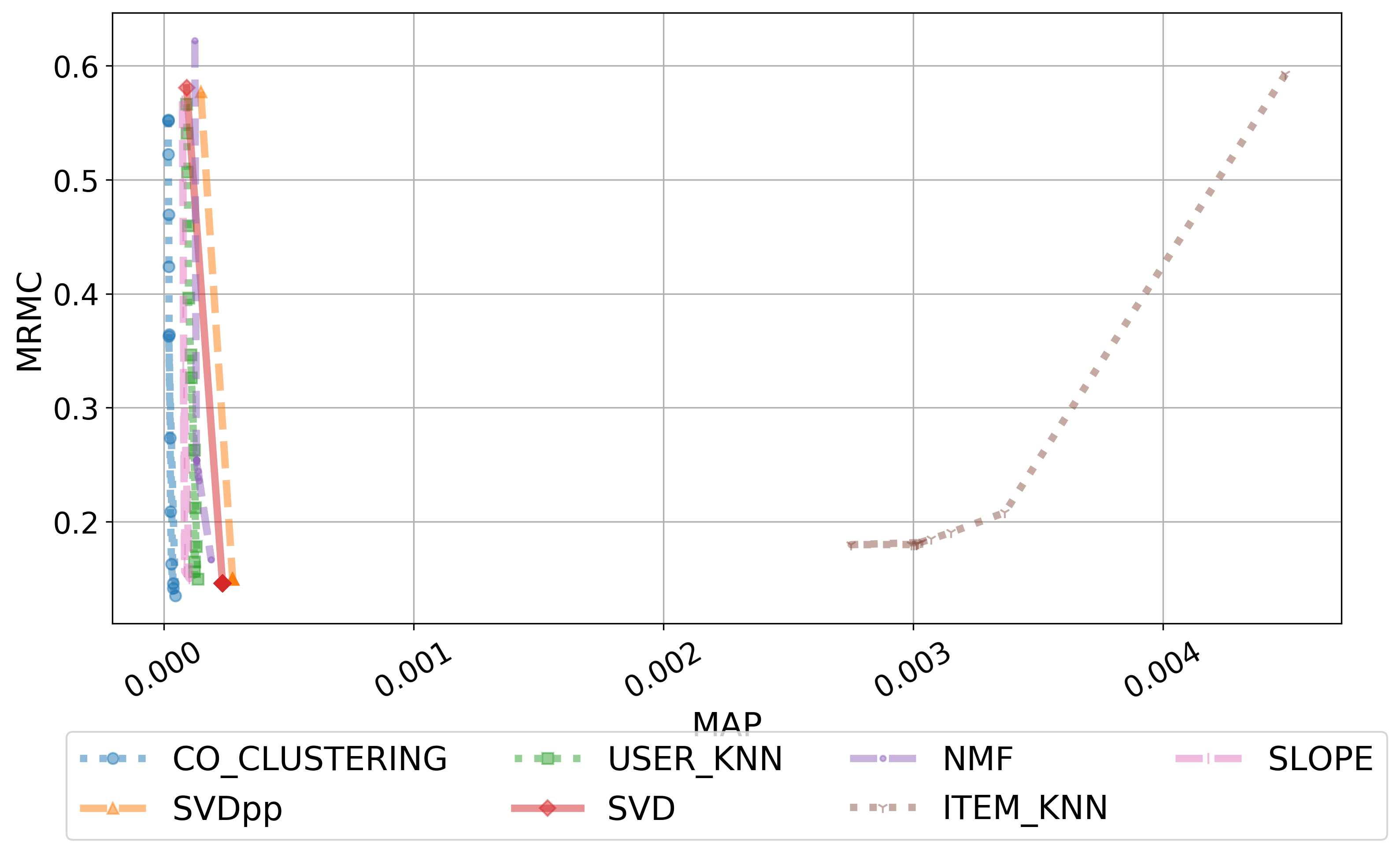}
		\caption{$\chi^2$ and $LIN$}
		\label{fig:oms_map_mrmc_chi_lin}
	\end{subfigure}
	\begin{subfigure}{0.48\textwidth}
		\centering
		\includegraphics[width=\linewidth]{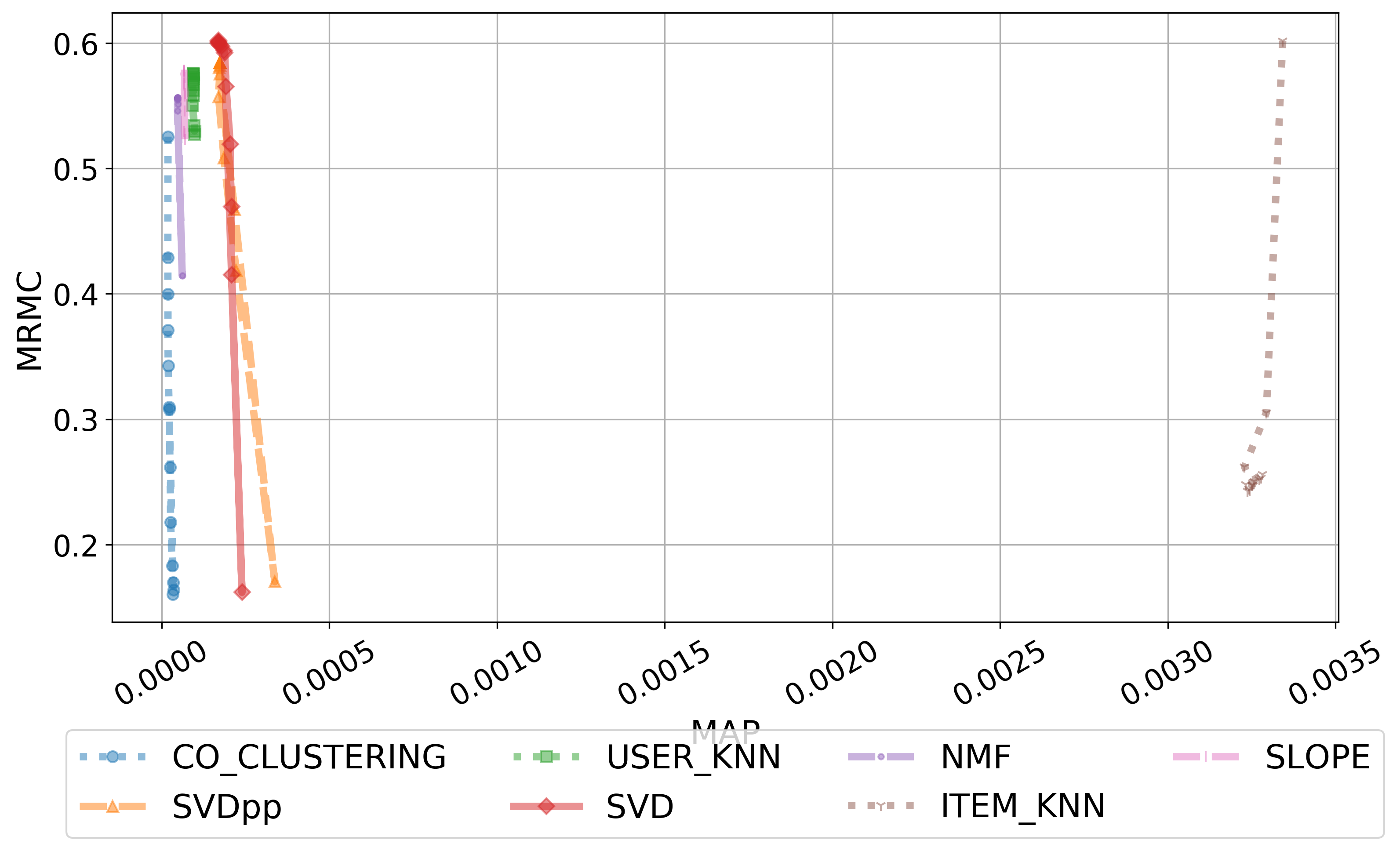}
		\caption{$\chi^2$ and $LOG$}
		\label{fig:oms_map_mrmc_chi_log}
	\end{subfigure}
	
	\caption{Taste Profile - MAP (x) and MRMC (y) - Results from $LIN$ and $LOG$ trade-offs combined with the KL-divergence, Hellinger and Pearson Chi Square.}
	\label{fig:oms_map_mrmc}
\end{figure*}

Another crossed-metric that we will analyze is the results from MAP and MRMC. Similar to the MAPxMACE, we want to understand the behavior of each system combination. Figure \ref{fig:ml_map_mrmc} presents the results from the Movielens, as well as Table \ref{tab:CMC_ML_MRMC_MAP} presents the CMC. Figure \ref{fig:oms_map_mrmc} presents the results from the Taste Profile, as well as Table \ref{tab:CMC_OMS_MRMC_MAP} presents the CMC values. 

\subsubsection{Movielens}
Similar to the MAPxMACE, we will start our analysis by the \textbf{KL-divergence} system combinations. Figures \ref{fig:ml_map_mrmc_kl_lin} and \ref{fig:ml_map_mrmc_kl_log} along with Table \ref{tab:CMC_ML_MRMC_MAP} (Lines 1 and 2) show the results from MAP and MRMC. It is possible to verify that the matrix factorization approaches obtained the best performance in the measure (KL). In Table \ref{tab:CMC_ML_MRMC_MAP} (Lines 1 and 2), we can observe that the KL-LOG-SVD++ obtained the best performance with a $CMC=3.05$. In the $LOG$ trade-off the second and third place is the SVD ($CMC=4.31$) and NMF ($CMC=5.76$) sequentially. In the $LIN$, the SVD++ obtained the best performance, whereas the combination KL-LIN-SVD++ ($CMC=3.52$) obtained the second best performance when we observe all the KL-divergence combinations. As we can verify that the MAPxMRMC and the MAPxMACE obtain the same recommender places to the KL-divergence.

For the \textbf{Hellinger} divergence measure, we will analyze Figures \ref{fig:ml_map_mrmc_he_lin} and \ref{fig:ml_map_mrmc_he_log} along with Table \ref{tab:CMC_ML_MRMC_MAP} (Lines 3 and 4). Similar to the KL-divergence, the matrix factorization approaches obtained all the best performances. However, the HE-LIN-NMF obtained the best performance to the measure with a $CMC=20.22$, followed by the HE-LIN-SVD++ with a $CMC=20.56$. To the $LOG$ trade-off, the best performance is HE-LOG-SVD++ with a $CMC=20.95$.

As the last analysis we will observe the \textbf{Pearson Chi Square} ($\chi^2$) combinations take into account Figures \ref{fig:ml_map_mrmc_chi_lin} and \ref{fig:ml_map_mrmc_chi_log} and Table \ref{tab:CMC_ML_MRMC_MAP} (Lines 5 and 6). Similar to the MAPxMACE results, as well as the previously presented measures, the matrix factorization approaches obtained all best performances in the $\chi^2$. However, in this measure the first place belongs to the SVD++ in the $LOG$ with a $CMC=9.08$ and in the $LIN$ with a $CMC=9.39$.

Based on the $CMC$ results we can observe that if the system is focused on MAP and MRMC and according to our experimental setup, the best combination to be implemented in the Movielens dataset is the KL-LOG-SVD++ with a $CMC=3.05$ as presented in the Table \ref{tab:CMC_ML_MRMC_MAP} Line 2. The second best performance belongs to KL-LIN-SVD++. When we observe the Figure \ref{fig:ml_map_mrmc}, it is possible to evidence that the metrics are not directly correlated. We note that the $LOG$ trade-off is part of this research proposes.

\begin{table*}[!t]
	\centering
	
	\caption{Taste Profile - CMC - Results from $LIN$ and $LOG$ trade-offs combined with the KL, HE and CHI divergence measure for each recommender algorithm.}
	\label{tab:CMC_OMS_MRMC_MAP}
	\resizebox{\textwidth}{!}{%
    \begin{tabular}{c|c|ccccccc}
    \hline
    \textbf{Divergence} & \textbf{Trade-off} & \textbf{\textbf{Co Clustering}} & \textbf{\textbf{Item KNN}} & \textbf{\textbf{NMF}} & \textbf{\textbf{Slope One}} & \textbf{\textbf{SVD}} & \textbf{\textbf{SVD++}} & \textbf{User KNN} \\ \hline
    KL                  & $LIN$              & 25124.25                        & {\ul \textbf{76.12}}       & 2299.61               & 6598.41                     & 953.96                & 624.99                  & 5331.51           \\
    KL                  & $LOG$              & 22937.04                        & {\ul \textbf{115.76}}      & 11416.07              & 8539.23                     & 3251.54               & 2971.04                 & 6144.5            \\ \hline
    HE                  & $LIN$              & 42028.72                        & {\ul \textbf{119.23}}      & 3775.94               & 11175.3                     & 1635.21               & 1452.92                 & 9181.55           \\
    HE                  & $LOG$              & 42531.68                        & {\ul \textbf{239.19}}      & 19002.44              & 13655.52                    & 5243.78               & 4729.24                 & 9891.66           \\ \hline
    CHI                 & $LIN$              & 12640.08                        & {\ul \textbf{68.46}}       & 1987.01               & 3937.01                     & 798.85                & 687.41                  & 2939.81           \\
    CHI                 & $LOG$              & 11985.59                        & {\ul \textbf{86.13}}       & 11219.71              & 8244.53                     & 2818.63               & 2696.87                 & 5937.93           \\ \hline
    \end{tabular}%
    }
\end{table*}

\subsubsection{Taste Profile}
Similar to the Movielens analysis, in the Taste Profile we will start our results by the \textbf{KL-divergence}. Figures \ref{fig:oms_map_mrmc_kl_lin} and \ref{fig:oms_map_mrmc_kl_log} and Table \ref{tab:CMC_OMS_MRMC_MAP} (Lines 1 and 2) show that the best recommender performance, in the KL-divergence, belongs to the Item-KNN both in $LIN$ with a $CMC=76.12$ and $LOG$ with a $CMC=115.76$ respectively. The second and third best recommender performances are the matrix factorization SVD++ and SVD. The Item-KNN system combinations obtains the best performance with a large difference to the second and third place. The worst performance in the KL belongs to the recommender Co Clustering.

For the \textbf{Hellinger} measure in the Taste Profile, Figures \ref{fig:oms_map_mrmc_he_lin} and \ref{fig:oms_map_mrmc_he_log} and Table \ref{tab:CMC_OMS_MRMC_MAP} (Lines 3 and 4) present the results to the measure. Similar to the KL-divergence, the best performance of the Hellinger is the Item-KNN with the $LIN$ trade-off, as well as the second and third places are the matrix factorization approaches. The Co Clustering approach achieves with the worst performance.

As to the \textbf{Pearson Chi Square} presented in Figures \ref{fig:oms_map_mrmc_chi_lin} and \ref{fig:oms_map_mrmc_chi_log} and Table \ref{tab:CMC_OMS_MRMC_MAP} (lines 5 and 6), similar to the previous measures and to the MAPxMACE results of the Taste Profile, the Item-KNN obtained the best performance to the $\chi^2$. The achieved values to the $LIN$ trade-off by the $CMC$ is $68.46$ and to the $LOG$ trade-off is $CMC=86.13$. The second and third place are the matrix factorizations, as well as the worst performance remains with the Co Clustering approach.

Based on the results we can observe that the CHI-LIN-Item-KNN is the best combination to be implemented with the Taste Profile. This analysis is based on the $CMC=68.46$. When we observe the table, it is possible to verify that the second best performance belongs to the KL-LIN-Item-KNN combination with $CMC=76.12$, followed by the combination CHI-LOG-Item-KNN.

\subsection{The Decision Protocol}
\label{sec:sub:decision}

The results presented so far help understanding the performance of each system combination. Nevertheless, only the CCE and CMC do not indicate the best combination to be implemented to a given domain. Based on this, we will apply the decision protocol to chose the best combination.

\begin{table*}[t!]
	\centering
	
	\caption{Protocol performance values to each system combination using the Movielens dataset.}
	\label{tab:decision_movielens}
	\resizebox{\textwidth}{!}{%
    \begin{tabular}{c|c|ccccccc}
    \hline
    \textbf{Divergence} & \textbf{Trade-off} & \textbf{\textbf{Co Clustering}} & \textbf{\textbf{Item KNN}} & \textbf{\textbf{NMF}} & \textbf{\textbf{Slope One}} & \textbf{\textbf{SVD}} & \textbf{SVD++}       & \textbf{User KNN} \\ \hline
    KL                  & LIN                & 244.1                           & 365.4                      & 16.7                  & 224.86                      & 19.83                 & 14.06                & 1682.22           \\
    KL                  & LOG                & 268.94                          & 484.4                      & 26.8                  & 321.3                       & 18.65                 & 13.14                & 575             \\ \hline
    HE                  & LIN                & 333.64                          & 511.86                     & 22.73                 & 318.35                      & 35.46                 & 22.94                & 6030.3            \\
    HE                  & LOG                & 634.17                          & 1811.66                    & 91.73                 & 1086.53                     & 34.07                 & 23.03                & 677.71            \\ \hline
    CHI                 & LIN                & \textbf{194.79}                 & \textbf{291.51}            & \textbf{13.44}        & \textbf{183.62}             & 17.5                  & {\ul 12.45}                & 1731.47           \\
    CHI                 & LOG                & 235.02                          & 382.99                     & 20.67                 & 252.74                      & \textbf{17.1}         & {\ul \textbf{12.14}} & \textbf{512.16}   \\ \hline
    \end{tabular}%
    }
\end{table*}

\subsubsection{Movielens}
Table \ref{tab:decision_movielens} presents the systems combinations performance values to the Movielens dataset. Based on the performance we can observe that the best combination is \textbf{CHI-LOG-SVD++} ($s = 12.14$). The Pearson Chi Square divergence obtained the best results due to its moderate performance in the CCE and in some results the best performance in the CMC. The $LOG$ trade-off balance, a part of our proposal, obtained the best performance in many results in the CCE and CMC evaluation. The SVD++ remains in the first position in the majority of the results of the Movielens. The Table \ref{tab:decision_movielens} presents in bold the best performance to each recommender. The CHI-LIN-SVD++ achieves the second best performance when all performances are compared. The worst performance are the KNNs.

\begin{table*}[t!]
	\centering
	
	\caption{Protocol performance values to each system combination using the Taste Profile dataset.}
	\label{tab:decision_oms}
	\resizebox{\textwidth}{!}{%
    \begin{tabular}{c|c|ccccccc}
    \hline
    \textbf{Divergence} & \textbf{Trade-off} & \textbf{\textbf{Co Clustering}} & \textbf{\textbf{Item KNN}} & \textbf{\textbf{NMF}} & \textbf{\textbf{Slope One}} & \textbf{SVD}     & \textbf{SVD++}  & \textbf{User KNN} \\ \hline
    KL                  & LIN                & 30217.86                        & 101.85                     & \textbf{2967.28}      & 8069.44                     & 1443.41          & \textbf{954.38} & 6507.13           \\
    KL                  & LOG                & 28136.9                         & 142.44                     & 13723.29              & 10253.62                    & 3973.18          & 3530.41         & 7376.52           \\ \hline
    HE                  & LIN                & 47101.47                        & 133.62                     & 4235.29               & 12581.75                    & 1939.16          & 1731.9          & 10329.72          \\
    HE                  & LOG                & 47589.66                        & 260.51                     & 21338.53              & 15365.32                    & 5953.06          & 5281.65         & 11126.32          \\ \hline
    CHI                 & LIN                & 16948.48                        & {\ul \textbf{91.75}}       & 2456.91               & \textbf{5375.44}            & \textbf{1107.25} & 970.79          & \textbf{3987.65}  \\
    CHI                 & LOG                & \textbf{16428.1}                & 113.77                     & 13507.54              & 9956.75                     & 3478.12          & 3228.84         & 7161.56           \\ \hline
    \end{tabular}%
    }
\end{table*}

\subsubsection{Taste Profile}
Table \ref{tab:decision_oms} presents the systems' combinations performance values to the Taste Profile dataset. According to the results, we observe that the best combination is \textbf{CHI-LIN-Item-KNN} ($s = 91.75$). The Pearson Chi Square divergence achieved the best results due to its best performance in the CCE and in the CMC. The $LIN$ trade-off balance, original \cite{Steck:2018} proposals, obtained the best performance in all results (first and second place). The Item-KNN remains in the first position in all results of the Taste Profile. Similar to the Movielens tests, the Table \ref{tab:decision_oms} presents in bold the best performance to each recommender and underlined the best performance. The worst performance belongs to the Co Clustering approach.

\subsection{Answering research questions}
\label{sec:sub:research_question}
In this section we debate each research question presented in Section \ref{intro}.

\textbf{RQ1}: \textit{How to find the best calibrated system to each domain?} As a research proposition we present a decision protocol, which helps us finding the best calibrated system to each used dataset.

\textbf{RQ2}: \textit{Do the calibrated systems have the same performance on the datasets? Are there differences among the systems behavior?} Among the results we can observe that each calibrated system has a different behavior and performance on the datasets, i. e., there are no best calibrated systems through the datasets, each dataset has its system that performs better. This affirmation enhances the necessity of the decision protocol, due to each dataset presenting different results.

\textbf{RQ3}: \textit{Is it possible to create a framework model for calibrated systems? What are the important components to these systems?} As we see along this research the proposed framework can be reproducible and provides an N-calibrated system. As well, the system components can be changed to search for the best performance on the domain dataset.

\textbf{RQ4}: \textit{When the user's bias is taken into account in the calibration trade-off, is it possible to improve the system performance?} As the Movielens and Taste Profile results show when we consider the users' bias in the formulation the performance is increased. This behavior is more observable in the Movielens results.

\subsection{Discussion}
\label{sec:sub:discussion}
The results show that the matrix factorization approaches are prominent recommenders when used with the calibration methods. In the Taste Profile, the Item-KNN obtains the best performance followed by the matrix factorization. As we can see in the results in some results the calibration method increases the precision as well as reduces the miscalibration.

The $LOG$ trade-off approach shows that it is possible to consider the users' bias in the equation and increase the overall performance. The framework can be easily applied and creates thousands of systems combinations. The right combination to be implemented can be chosen by the proposed protocol, thus facilitating the exploitation of optimal performances.

The claim made by \cite{Steck:2018, Kaya:2019} about the drop of precision can be partially evidenced from our results. Other colleagues claim the same, in which the Pearson achieves competitive results such as \cite{DASILVA2021115112}. Likewise \cite{DASILVA2021115112}, our findings point towards the possibility to implement and test new divergence measures in the calibrated recommendations.
\section{Conclusion and Future Work}
\label{sec:conclusion}
In this research, we exploited the calibrated recommendations in a user-centric view. We show that the state-of-the-art does not consider the users' bias in the trade-off balance. Based on this premise, we propose the $LOG$ trade-off, showing that it is possible to increase the MAP, MACE and MRMC performance. We also showed that the state-of-the-art explores different numbers of implementations without an automatic way to decide the best implementation. In order to tackle this problem, we propose two coefficients and a decision protocol that indicates the best implementation to the dataset. In addition, we showed the possibilities of implementing the calibrated system following the framework division in components to provide an easier way to be implemented and tested. 

As a future work, we plan to investigate other calibration measures and compare the effect of the post-processing in the collaborative filtering and content-based filtering. Another extension is to explore different types of distribution, for instance, compare the normalized against the non-normalized ones. In addition, we will investigate other relevance measures.

\begin{acks}
This study was financed by the Coordenação de Aperfeiçoamento de Pessoal de Nível Superior - Brasil (CAPES) - Finance Code 88887.502736 / 2020-00. The authors acknowledge the National Laboratory for Scientific Computing (LNCC/MCTI, Brazil) for providing HPC resources of the SDumont supercomputer, which has contributed to the research results reported within this paper. URL: http://sdumont.lncc.br
\end{acks}

\bibliographystyle{ACM-Reference-Format}
\bibliography{acmart}










\end{document}